\documentclass[paper,12pt]{JHEP3}
\usepackage{amssymb}

\renewcommand{\b}{\beta}
\newcommand{\g}{\gamma}     \newcommand{\G}{\Gamma}

\renewcommand{\l}{\lambda}  
\newcommand{\bl}{\bar\lambda}

\renewcommand{\G}{\Gamma}

\renewcommand{\r}{\rho}
\newcommand{\brho}{\bar\rho}
\newcommand{\m}{\mu}

\newcommand{\U}{\Upsilon}
\newcommand{\h}{H}
\newcommand{\dr}{\partial}

\newcommand{\bz}{\bar{z}}
\newcommand{\ba}{\bar{a}}
\newcommand{\LR}{L|R}

\newcommand{\bgg}{\bar\gamma}

\newcommand{\be}{\begin{equation}}
\newcommand{\ee}{\end{equation}}
\newcommand{\bea}{\begin{eqnarray}}
\newcommand{\eea}{\end{eqnarray}}
\newcommand{\bary}{\begin{array}}
\newcommand{\eary}{\end{array}}

\newcommand{\nn} {\nonumber \\}
\newcommand{\spa}{\hspace{1cm}}

\newcommand {\eqr} [1]  {{(\ref{#1})}}

\newcommand{\file}[1]{}
\newcommand{\tila}{{\tilde a}}
\newcommand{\btila} {\bar\tila}

\newcommand{\cC}    {{\mathcal{C}}}

\newcommand{\cH}    {{\mathcal{H}}}

\newcommand{\cI}    {{\mathcal I}}
\newcommand{\cJ}    {{\mathcal J}}
\newcommand{\cL}    {{\mathcal{L}}}

\newcommand{\cP}    {{\mathcal{P}}}

\newcommand{\cT}    {{\mathcal T}}

\newcommand{\hV}  {{\hat V}}
\newcommand{\tilV}  {{\widetilde V}}

\newcommand{\vtil}  {{\tilde v}}

\newcommand{\half}  {\frac 1 2}

\newcommand{\pd}    {\partial}
\newcommand{\bpd}      {{\bar \partial}}
\newcommand{\bra}{\langle}
\newcommand{\ket}{\rangle}

\preprint{CALT-68-2438\\hep-th/0309005}
\title{Strings in Gravimagnetic Fields}

\author{
Yeuk-Kwan E. Cheung$^{1,2}$, Laurent Freidel$^{2,3}$ and Konstantin Savvidy$^2$\footnote{
\tt cheung@theory.caltech.edu, lfreidel, ksavvidis@perimeterinstitute.ca}\\ \\
$^1$California Institute of Technology, Theory Group M.C.
452-48\\  \, \,  Pasadena CA 91125, U.S.A.\\  \\
$^2$Perimeter Institute for Theoretical Physics, 35 King St North, \\
 \, \, Waterloo, ON N2J  2W9, Canada\\  \\
$^3$Laboratoire de Physique, \'Ecole Normale Sup{\'e}rieure de Lyon \\
\, \, 46 all{\'e}e d'Italie, 69364 Lyon Cedex 07, France
}

\abstract{
We provide a complete solution of  closed strings propagating in Nappi-Witten  space.
Based on the analysis of geodesics we construct the coherent
wavefunctions which approximate as closely as possible the classical trajectories.
We then present a new free field realization of the current algebra using
 the $\gamma, \beta$  ghost system.
Finally we construct the quantum vertex operators, for the  tachyon, by representing
the wavefunctions in terms of the free fields.
This allows us to compute the three- and four-point amplitudes,  and propose the
general result for N-point tachyon scattering amplitude.
}

\begin{document}

\section{Introduction}

Understanding string dynamics in previously inaccessible supergravity
backgrounds has been made possible by the realization that on the
maximally symmetric backgrounds of PP-waves string theory becomes
exactly soluble, even in the presence of Ramond-Ramond fields
\cite{metsaev, Blau:2001ne, Berenstein:2002jq}.  Also
it has been known that certain backgrounds with constant field
strengths of Neveu-Schwarz B-field are soluble in the same PP-wave
limit \cite{Russo:2002rq}.

There has been a flurry of studies of strings in pp-wave backgrounds
prompted by the realization that such models become free in the light-cone gauge.
This approach has been successful in understanding the spectrum of the
theory.  However the light-cone
formalism prevents us from dealing with string interactions
using familiar worldsheet techniques for computing
scattering amplitudes, forcing upon us the light-cone
string field theory approach \cite{ppstringFT}.
The main purpose of this paper is to  describe
string interactions in a pp-wave background
in a fully covariant formalism where the worldsheet theory is interacting.
We shall only address the question in the context of a pp-wave
background with NS-NS flux but no R-R flux.  
We should  mention that this issue
has also been taken up recently by G.~D'Appollonio and E.~Kiritsis~\cite{D'Appollonio:2003dr}.
However the approach in our paper is completely different and it allows
us to compute any N-point scattering amplitudes, including but not limited to three-
and four-points.

The specific model which we consider is the Nappi-Witten model \cite{Nappi:1993ie}.
It does as closely as possible represent the situation of
flat space with NS flux, by having metric deformed
away from flat space to take into account the curvature introduced by
the flux. There is unfortunately no limit available where one may be
able to neglect the curvature, and consider the torsion field alone.
Thus in order to understand the effect of torsion it will be necessary
to disentangle the curvature from torsion \emph{a posteriori}.
Certainly it is true that the shape
of closed strings is distorted by the curvature and torsion.
For  large values of the torsion the usual vacua  are unstable and  new vacua,
the ``long-string states' of~\cite{Seiberg:1999xz}
carrying  winding numbers are created.
We discuss this dynamical dielectric-type effect in Section~\ref{sec:lc}.

Though formally similar to the $SL(2,R)$ and the $SU(2)$ WZW models,
the Nappi-Witten model  is in many ways simpler.
For example the naive value $c=4$ for the central charge holds \cite{Nappi:1993ie},
it is equal to the dimension of the group.
Considerable simplification is due to the particularly simple form
of the cubic interactions on the worldsheet as well as the fact that
 Nappi-Witten background being an exact solution of the string sigma model.
The main technical reason which allows us to completely solve the
theory in this case is the fact that we have found a mapping of the Nappi-Witten
model to a free field theory.  We can then identify  the symmetry currents and
construct the vertex operators in terms of the free  fields.

We should emphasize that our approach is very different from the earlier
attempts \cite{Kiritsis:jk, kkl, kirpio} to solve the NW model in terms of
quasi-free (or twist) fields which do not allow an explicit construction of the
vertex operators. Our proposal for the N-point tachyon amplitude is new.
So is a relation between amplitudes having a different number of
conjugate fields.
Our results for the three and four points function agree with
the ones in~\cite{D'Appollonio:2003dr}.

We have tried in this paper to give a complete and consistent
description of closed strings  propagating in Nappi-Witten space.
We review in Section~\ref{sec:background}
 the classical picture of string propagation in the light-cone gauge
 and in Section~\ref{sec:wave}  the  semi-classical wave propagation. 
The classical analysis gives us
a nice physical picture and provides us with the geometric intuition
to guide us in the covariant (algebraic) approach.

The Wakimoto free field realization of the algebra, using $\beta, \gamma$ ``ghosts''
is  introduced in Section~\ref{sec:gammabeta}.
This  is formally similar to the one
 used for $SL(2,R)$ and $SU(2)$.
We are able to give  a 
 geometrical interpretation of these subsidiary fields in
terms of the original coordinate fields not only for  $\gamma$,
which is a simple coordinate redefinition, but also for the $\beta$  field.
We then construct  the vertex operators in terms of these fields and 
the conjugate vertex operators via an integral transformation.
These constructions are presented in
Section~\ref{sec:vertexop}.
The extra simplification, as compared to $SL(2,R)$ and $SU(2)$ cases,
is due to the fact that the
screening charge contains only null fields and thus producing no
contractions when we perform the vertex operator correlator
calculation (in Section~\ref{sec:correlation}).
The final expression for the $N$-point amplitude with one conjugate
vertex operator is, instead of a complicated multiple 
integral of the Fateev-Dotsenko type,
a single surface integral in  just one complex variable.
We present some checks on this amplitude in Section~\ref{sec:ward} using
conformal and algebraic Ward identities as well as the Knizhnik-Zamolodchikov
equation. The N-point amplitude with more than one conjugate field is
given in term of an integral transform of the original amplitude.
We also present explicit justifications for this rule in the case of three and
four point amplitudes.
The ``long strings'' appear  already as poles in the three-point amplitudes.
The four-point  amplitudes and the factorization property are  analyzed.
Finally we discuss
the flat space limit of our general N-point amplitudes in Section~\ref{sec:flatspace}.

Several appendices contain detailed derivations which serve to make our 
paper self-contained as well as to correct some misprints in the literature.
Appendix A contains a thorough analysis in the light cone gauge.  Appendix B reviews 
the representation theory of the Nappi-Witten algebra and  
its link with the wave functionals on the group manifold.  
In Appendix C we derive the integral transform of the conjugate vertex operators.  Finally 
Appendix D details proof of the chiral splitting formulae used in the computation of the 
correlation functions.

\section{Strings on Nappi-Witten spacetime} 
\label{sec:background}

The Nappi-Witten model~\cite{Witten:ar} is a WZW  model on 
the centrally-extended two-dimensional Poincare group. 
The solvability of string theory in this group manifold(the Nappi-Witten space) 
relies on this  underlying infinite dimensional symmetry\footnote{The existence of 
infinite dimensional symmetry may also play a role in solving string
theory on {$\mathrm{AdS_5\times S^5}$} \cite{Bena:2003wd, Dolan:2003uh}.}.
$J_\pm, J, T$ are the anti-hermitian generators of the algebra:
\be \label{algebra}
[J^{+}, \,J^{-} ] =2\, i \, T ~,~~~
[J, \,J^{+}] = i\, J^{+} ~,~~~ [J,\, J^{-} ]=-i\,
J^{-} ~,~~~ [T, \,\diamondsuit] = 0~.
\ee
A generic group element is given by
\be \label{group}
g(a,u,v)= e^{\h a J^+ +\h \ba J^-} \, e^{\h uJ +\h vT}~,
\ee
where $a=\half (a^1+ i \, a^2)$. The group properties are spelled out in 
Appendix~\ref{representation}.

The Nappi-Witten space is
the group manifold, parametrized by  $a,\bar{a},u,v$, whose metric is given by
\bea \label{NWmetric}
ds^2 &=&  \frac{1}{\h^2} \mathrm{tr} (g^{-1} \, dg )^2   \nn
&=& 2\, du\, dv + 4da\, d\ba + 2i\,(\ba \, da- a\, d\ba)\, du~.
\eea
In addition there is a NS B-field with constant field strength H:
\be
B_{ui} = -H \, \epsilon_{ij} \, a^j \spa H_{u12} = -H~.
\ee

A simple analysis 
shows that the coordinate
direction corresponding to $J$ is a null direction which should be
identified with the $u=t+\psi$ of the pp-wave limit.  The fact that
translations along the light cone time do not commute with spatial
translations but instead  generate rotations in the spatial plane
should be interpreted as angular momentum carried by the
circularly polarized plane wave.

The metric~\eqr{NWmetric} is invariant under the isometry group $G_L \times G_R$,
$g\rightarrow g_L^{-1} \, g \, g_R$.
 Infinitesimally the  group action  is given by
\be\label{conscharge}
\begin{array}{rclrcl}
T_L & = & -\dr_v~, &
T_R & = & \dr_v~,\\
J_L &=& -(\dr_u + i \, (a \, \dr_a -\bar{a} \, \dr_{\bar{a}}))~,~~~ &
J_R &=& \dr_u~, \\
J_L^+ & = & -(\dr_{a}  + i \h \ba \, \dr_v)~,~~~ &
J_R^+ &= & e^{i\h u} (\dr_{a} - i \h\ba  \, \dr_v )~,\\
J_L^- &=& -(\dr_{\ba}  -i \h a \, \dr_v)~,    &
J_R^-&=& e^{-i\h u} (\dr_{\ba} +i \h a  \, \dr_v)~.
\end{array}
\ee
The generator $T$ generates translations in $v$,
$J_R$ generates translations in the $u$ direction,
and $J_L+J_R$ rotations in the transverse plane. The others
generate some twisted translations in the transverse plane. 
Overall, the isometry group is 7-dimensional\footnote{The  commuting generator $T$ 
should be counted only once.} and consists of
two commuting copies  of the Nappi-Witten algebra.
Remarkably  boosts are not among these symmetries
even in the limit of flat space $H \rightarrow 0$. This
makes it impossible to use old techniques of setting
the light-cone momenta to zero and later using the symmetries
to recover the fully general results.

Some remarks are due on this nonabelian group of isometry.
First  it is impossible to diagonalize all the generators at once. 
Second, the mutually commuting but inequivalent action of the descendants of the other
chiral copy is to be taken into account. The net result is that a
state is characterized by four quantum numbers: two real momenta
$p^+, p^-$ and two complex numbers  $\rho$ and $\lambda$  which parametrize
the position and the radius of the corresponding classical trajectory.
One may be surprised that the position is a quantum number characterizing   a state. 
But as we shall see,  $\lambda$ and $\rho$ are indeed the
eigenvalues of some of the isometry generators: they are  the
zero mode part of the current algebra.

Finally the corresponding string sigma model action reads\footnote{We are 
ignoring the other six directions.   We have also omitted the fermions.}
\bea 
\label{NWsigmamodel}
S(g) = \frac{1}{4\pi}  \int d\sigma^2  
&\{& \sqrt{-h} \, h^{\alpha\beta}\, [\, 2\, \pd_\alpha u\, \pd_\beta v + \pd_\alpha a^i \,\pd_\beta a^i
  -H\, (a^1\,\pd_\alpha a^2 - a^2 \,\pd_\alpha a^1)\, \pd_\beta u \, ]\nn
&& -H \,\epsilon^{\alpha\beta}\, (a^1\,\pd_{\alpha} a^2 -a^2\,\pd_{\alpha}a^1)\,\pd_\beta \, u
~\} ~,
\eea
where $H$ is related to the level $k$ of the WZW model by $k=H^{-2}$.
After  conformal gauge-fixing  the terms due to
curvature and torsion can be naturally combined:
\be
S(g) = \frac{1}{4\pi} \int 
2 \, \pd_{+} \, u \pd_{-} v +  \pd_{+} a^i \, \pd_{-} a^i -
H \, ( a^1 \, \pd_- a^{2} - a^{2} \, \pd_{-}a^{1} )\,  \pd_+ u ~.
\ee
And hence  our observation that the effects of curvature and torsion are inseparable,
in the sense that there is no (obvious) way to take a limit in which the metric
perturbation would go away while retaining the torsion.
Note that if we rescale $u$ by $H$ and $v$ by $H^{-1}$ we can absorb the
dependance in the coupling constant. We shall therefore suppose for  most 
parts of the paper  that   $H=1$.  
We shall reinstate the $H$ dependence at the very end when we discuss the flat
space limits of our N-point amplitude.

\subsection{Geodesic Motion}
\label{sec:geodesic}

A lot of intuition about the behaviour of strings can be gained
by first looking at the behaviour of their centre-of-mass coordinates,
\emph{i.e.} the classical picture of particles propagating in the
given geometry. 

The geodesic,  followed  from the metric (\ref{NWmetric}), is described by:
\bea
	u &=& u_0 + p^+ \, \tau   \nn
     a  &=&  -\rho  +  \lambda \, e^{+ip^+\tau} ~,\nn
\bar{a} &=& -\bar{\rho} + \bar{\lambda} \, e^{ -ip^+\tau} ~.
\eea
As advertised above, the integration constants   $\rho, \lambda$ represent 
the position and radius  of the circular trajectory.
Having found $a,\bar{a}$  we can now solve the remaining equation for $v$
\be
\label{eq:vgeo}
v = v_0 + (p^- + 2 \,  \, p^+ \, |\lambda|^2) \,  \tau  +
i(\bar{\lambda} \, {\rho} \, e^{-i   p^+\tau}-\lambda \, \bar{\rho} \, e^{+i   p^+\tau})
\ee
Lastly, if we are interested in massless particles, the condition for
the world-line to be light-like is
\[
2 \, p^+ \, p^-   +   \, |2 p^+\lambda|^2  = 0.
\]
As seen from this identity we can interpret $2 p^+\lambda$ to be the
momentum of transverse coordinates, $p_{\bot}$. 
This gives the parameter $\lambda$ a dual role:  on the one hand it is the radius
 of the transverse circle, on the other hand it  is
the transverse momentum  in units of $p^+$.
We shall see  later in Subsection~\ref{sec:lc} that at the quantum level this fact will manifest
itself as noncommutativity in  the transverse space.

Closed strings having  sufficiently
small light cone momenta do follow geodesics as expected on general grounds.
 The reason is that in deriving the geodesic equation
from string equation of motion one assumes the ground state to be $\sigma$-independent.
This is consistent with the closed string periodic boundary conditions.
However if the light cone momentum becomes big, the $\sigma$-independent configuration 
is not necessarily a minimum energy state.  
New vacua corresponding to long-string configurations, which have nontrivial $\sigma$ dependence, 
emerge.  So the effect of having a torsion field extending in the time direction leads 
to dynamical effects under which the geodesic is no longer the natural motion of the closed string.

For open strings the torsion has  even more drastic effects.
It affects the centre-of-mass motion of the open string and hence its
motion deviates from geodesics.  
The  two ends of the open string couple directly to $B+F$ with opposite
signs and the string becomes polarized as a dipole.
When $B+F$ is non-uniform the open string will experience a
net force proportional to the gradient of the field.  The situation is
furthermore complicated by the fact that the resulting physical description depends on 
the electromagnetic field, $F$, in the background.  

\subsection{Light Cone Analysis}
\label{sec:lc}

We would now go to the  light-cone gauge in which the cubic interaction term becomes quadratic
and the model becomes soluble.  
The light-cone Hamiltonian, $\cH_{lc}$, is
\bea    \label{eq:lightconeH}
\cH_{lc} &=& \half \Pi_1^2 + \half \Pi_2^2 + \, \mu \,(a_1 \Pi_2 -a_2 \Pi_1)
   + \half \acute{a}_1^{ 2} +\half \acute{a}_2^{ 2} +\, \mu \,  (\acute{a}_1 a_2 - \acute{a}_2 a_1)
   +\half \mu^2 \, ( a_1^{2} +a_2^{2} ). \nn
   &{}&
\eea
We can also solve for the longitudinal coordinate $\acute{v}$
in terms of the dynamical transverse physical fields:
\bea \label{eq:vprime}
\acute{v} &=& -(\acute{a}_1 \Pi_1 + \acute{a}_2 \Pi_2)  \nn
          &=&  - \dot{a_1} a_1^{\prime} - \dot{a_2} a_2^{\prime} -
                                \mu\, ( a_2 a_1^{\prime} - a_1 a_2^{\prime} )~.
\eea
The form of this last expression cannot be naively guessed and differs
from the flat-space result by the H-dependent terms.  This constraint,
integrated over $\sigma$, produces the left-right matching
condition on the physical Hilbert space.


The solutions to the equation of motion are given by
\bea    \label{modeexpansion}
a &= (a^1 + i \, a^2)/2
  &= i\, e^{i\mu\sigma^+} \left( \sum_{n\in\mathbb{Z}}
         \frac{\tilde{a}_{n} }{n+\mu} \, e^{-i(n+\mu)\sigma^+}
                     +  \frac{a_{n} }{n-\mu} \, e^{-i(n-\mu)\sigma^-}  \right)
\eea
\bea
\bar{a} &= (a^1 - i \, a^2)/2
  &= i\, e^{-i\mu\sigma^+} \left( \sum_{n\in\mathbb{Z}}
         \frac{\bar{\tilde{a}}_{n}}{n-\mu} \, e^{-i(n-\mu)\sigma^+}
                     +  \frac{\bar{a}_{n}}{n+\mu} \, e^{-i(n+\mu)\sigma^-}  \right)
\eea
where we have introduced the notations, $\mu \equiv p^+ H$, and $\sigma^{\pm}=\tau \pm \sigma$.

The $\tila_0$ ($\btila_0$) is the center-of-mass coordinates
and should be identified with $\rho$ ($\bar\rho$).
Similarly we should  identify $a_0$ ($\bar{a}_0$) with the radius, 
$\lambda$ ($\bar\lambda$), of the classical trajectory.
One  also observes  that the terms linear in $\tau$ are not allowed 
and thus there is no zero-mode momentum operators in the mode expansion above. 
If one takes the limit $H \rightarrow 0$, the frequency of the $a_0$ 
mode goes to zero and becomes the momentum operator of the limiting flat space.

\subsubsection*{Quantization}
\label{quantization}

%
Upon quantizing we obtain the commutation relations for the oscillators:
\bea 
	[ a_n , \ba_m ]  &=&   \half \delta_{n,-m}(n-\m),       \nn
 	{[} \tila_{n} , \btila_m {]}  &=& \half \delta_{n,-m}(n+\m).  
\eea
The reality condition gives
\be
	a_n^\dagger = \ba_{-n};   \,\,\,   \tila_n^\dagger =\btila_{-n}.
\ee

When  the value of $\mu$ is restricted to be $0\leq \mu<1$
creation operators are those with negative indices, $m<0$.
Unlike the case in flat space case, the right
and left moving zero-modes here are not degenerate, 
{\it i.e.} they have frequencies of $+\mu$ and $-\mu$ respectively.
This is reflected in the mode expansion already -- the right and 
left moving zero-modes are independent of each other.

When $\mu=N+\epsilon$ where $N$ is a positive integer the vacuum  is annihilated by
$ a_{N+m}, \bar{\tilde{a}}_{N+m}, \, m>0$ and
$ \tilde{a}_{-N+m}, \bar{{a}}_{-N+m},\, m\geq 0$.
The ``zero modes" are given by $ a_{N}, \tilde{a}_{-N}$ and the corresponding classical
 classical solution is
\be
\frac{ie^{i N\sigma^+}}{\epsilon}
    \left(\tilde{a}_{-N} -a_{N}e^{2i\epsilon\tau} \right).
\ee
When $N=0$ this is the geodesic motion,
 centering at $\tila_0/\mu$ and with a radius $ a_0/\mu$,
oscillating in time at frequency $\mu$.
 But for $N\neq 0$ this describes the motion
of a ``long string,"~\footnote{
See \cite{Seiberg:1999xz, maloog1} for a general definition and
\cite{kirpio, D'Appollonio:2003dr} for an application of this notion to the Nappi-Witten model.}

\be \label{longstring}
 e^{iN\sigma} \left( \frac{\tila_{-N}}{\epsilon} e^{iN\tau} 
 - \frac{a_N}{\epsilon}  e^{iN\tau} e^{2i\epsilon\tau} \right)
\ee
\\ 
{\it i.e.} 
a $2$-dimensional surface winding $N$ times around the origin.
It envelopes the geodesic -- centered at $\tila_{-N}/\epsilon$ and with 
a radius of $a_N/\epsilon$ -- 
and oscillates with a slow frequency $\epsilon$.
What we are witnessing here is a dynamical dielectric effect\cite{Myers:1999ps}
 such that the light-cone momentum is transmuted into winding number under
  the influence of the $H$ field:
For every increase of the light-cone momentum by unit value of 
$\frac{1}{2\pi \alpha\prime^2}H^{-1}$ 
the winding number of the ground state also increases by one.
The $\mu=0$ case, which correctly reproduce the flat space linear
motion, has been discussed above in the previous subsection.

Finally the quantum Hamiltonian can be obtained by substituting the string mode expansions
into the classical expression and normal-ordering:
\be                    \label{eq:ham}
\mathcal{H}_{lc} = \sum_{n \in \mathbb{Z}} \left(
:a_{n} \bar{a}_{-n}: \, \frac{n}{n-\mu}+
:\tilde{a}_{n}  \btila_{-n}: \, \frac{n+2\mu}{n+\mu}\right)
\ee

\subsection{Wavefunctions} 
\label{sec:wave}

In this section we follow on to the semi-classical analysis with
the construction of the  wave functions on the group manifold.
The classical picture will again provide the guiding principle: we
will choose the wavefunctions to approximate the classical trajectories 
as close as possible. 
As a result the Landau-like orbits will translate into 
the coherent wavefunctions in quantum mechanics.  
This exercise provides us with  more intuition in the propagation and
interaction of particles on the Nappi-Witten space. 
It will also be our starting
point for writing down the vertex operator in Section~\ref{sec:vertex}. 
App.~\ref{representation} 
contains the  group theoretical analysis and  the representation theory of the 
Nappi-Witten algebra. Readers who are not familiar with these aspects of the
model should consult it at this point.

The quantum wavefunctions  are eigenfunctions  of the covariant Laplacian operator
on the group manifold:
\be
\label{eq:laplacian}
\Delta = \frac{\pd^2}{\pd a_1^2} +
\frac{\pd^2}{\pd a_2^2} +
2 \, \frac{\pd}{\pd u} \, \frac{\pd}{\pd v} +
\frac{1}{4} {(a_1^2 + a_2^2)}\, \frac{\pd^2}{\pd v^2} +
 \left( a_1\, \frac{\pd}{\pd a_2} - a_2\,
\frac{\pd}{\pd a_1} \right) \, \frac{\pd}{\pd v}~.
\ee
The Laplacian is equal to the Casimir for the left or right symmetry generators 
(\ref{conscharge})
\bea
\Delta &=& 2 \, J_L T_L + (J^+_LJ^-_L +J^-_LJ^+_L)/2  \nn
&=&  2 \, J_R T_R + (J^+_RJ^-_R +J^-_RJ^+_R)/2~.
\eea
This implies that the eigenfunctions of the Laplacian with eigenvalues
\be
\cC= -2 \,  p^+ \, (p^- + \half)
\ee
 are given by the matrix elements of $g$ in
the representations $V^{p^+,p^-}$ 
(or in the conjugate representation, $\widetilde{V}^{p^+,p^-}$). 
In our convention (see Appendix B.1) $V^{p^+,p^-}$ representation
has the highest weight with respect to $J$, whereas $\widetilde{V}^{p^+,p^-}$ has the lowest weight.
They share the same Casimir given by  $\cC$.
Let us  denote matrix element of $g$ in the coherent state basis of $V^{p^+,p^-}$ 
by\footnote{Note that the wavefunction obtained this way is not  normalized. 
}
\be	\label{defwave}
	\phi^{p^+,p^-}_{\bar{\rho},\lambda}(g)= {\bra\rho|g|\lambda\ket}~.
\ee

Using the definition of the group element \eqr{group} and that of the coherent states
we get
\be	\label{wavefunc}
	\phi^{p^+,p^-}_{\bar{\rho},\lambda}(g)  = e^{ip^+v +ip^-u}  e^{-p^+ a\bar{a}}
	\exp\left[2 p^+(a\lambda e^{-iu} -\bar{\rho} \bar{a}) + 2 p^+ \bar{\rho} \lambda e^{-iu}\right]~,
\ee
The ground state wave functional is a plane wave state positioned at $a =0$ in the transverse plane
\be
	\phi^{p^+,p^-}_{0,0}(g)  = e^{ip^+v +ip^-u} e^{-p^+ a\bar{a}}~.
\ee

Normalized and written in a more suggestive way  
\be
\phi^{p^+,p^-}_{\bar{\rho},\lambda}(a,u,v)  
=e^{ip^+v +ip^-u} 
            e^{-p^+|\bar{a}-\lambda e^{-iu} +\rho|^2}  e^{p^+a (\rho + \lambda e^{-iu}) - c.c.} 
	 e^{- p^+( \bar{\lambda}\rho e^{iu} - c.c.)}.
\ee
the wave functional takes the form of  a Gaussian centered around
$\bar{a} = -\rho+\lambda e^{-iu}$. 
Moreover, it is a plane wave in $v$ of momentum $p^+$; and the semi-classical
momentum  in the transverse plane, $p^a$, is given by $\rho +\lambda e^{iu}$.
The direction $u$ has a more complicated semi-classical momentum:
$p^u = p^- - \lambda(a+\bar{\rho})e^{-iu} +c.c$.
All these results agree  with  the geodesics analysis in Section~\ref{sec:geodesic}.

Since the representation $\widetilde{V}^{p^+,p^-}$ is conjugate to $V^{p^+,p^-}$,
 the conjugate wave functional is related to the original wave functional by 
 complex conjugation
\be
\widetilde{\phi}^{p^+,p^-}_{\bar{\rho},\lambda}(g)=
\overline{{\phi}^{p^+,p^-}_{{\rho},\bar{\lambda}}(g)}~.
\ee
Explicitly this reads
\be \label{dualwave}
\widetilde{\phi}^{p^+,p^-}_{\bar{\rho},\lambda}(g) \nn
= e^{-ip^+v - ip^-u} e^{-p^+ a\bar{a}} e^{-p^+ \l\bl} e^{-p^+ \r\brho}
\exp\left[2 \,  p^+(\bar{a}\lambda e^{+iu} -\bar{\rho}{a} +\bar{\rho} \lambda e^{+iu})\right]~. \\
\ee
This corresponds to a wave moving backward in the  light-cone time  and centred around 
$a= \lambda e^{iu} -{\rho}$.
Finally  we can  construct the wave functions corresponding to the $p^+=0$ 
representation~(See App. \ref{representation}).

The wave functions we have constructed are eigenvectors under the action of the lowering isometry generators
$J^\pm_{\LR}$ (\ref{conscharge}):
\bea
J_L^- \phi^{p^+,p^-}_{\bar{\rho},\lambda}(g) =  (2 \, p^+\bar{\rho}) \phi^{p^+,p^-}_{\bar{\rho},\lambda}(g)~,&
\spa
J_R^+ \phi^{p^+,p^-}_{\bar{\rho},\lambda}(g) =  (2 \, p^+{\lambda}) \phi^{p^+,p^-}_{\bar{\rho},\lambda}(g)~,\nn
J_L^+ \widetilde{\phi}^{p^+,p^-}_{\bar{\rho},\lambda}(g) =  (2 \, p^+\bar{\rho}) \widetilde{\phi}^{p^+,p^-}_{\bar{\rho},\lambda}(g)~,
& \spa
J_R^- \widetilde{\phi}^{p^+,p^-}_{\bar{\rho},\lambda}(g) =  (2 \, p^+{\lambda}) \widetilde{\phi}^{p^+,p^-}_{\bar{\rho},\lambda}(g)~.
\eea
Together $\phi^{p^+,p^-}_{\bar{\rho},\lambda}$, $\widetilde{\phi}^{p^+,p^-}_{\bar{\rho},\lambda}$ 
form a complete basis of
normalisable ($L^2$ integrable) solutions of the wave equations\footnote{See Appendix~\ref{representation}.}
\be\label{waveeq}
\Delta \phi = -2 \, p^+(p^- +1/2)\phi~, ~~~  \partial_v \phi = \pm ip^+ \phi~~\textrm{where}~p^+>0~.
\ee

It will be very important for us to note that $\phi^{-p^+,-(p^-+1)}_{\bar{\rho},\lambda}(g)$
is also solution of (\ref{waveeq}).
This solution is well defined as a function on the group, however
it contains a factor   $\exp(2p^+ a\bar{a})$ which is unbounded.
It is hence not normalisable and cannot be taken as a wave functional corresponding
to the representation  $\widetilde{V}^{p^+,p^-}$.
However one can perform integral transform of this solution
\be\label{duality}
\int d^2\l ~e^{2p^+\l \rho} \, e^{2p^+\bar{\l} \bar{\rho}} \, {\phi}^{-p^+,-p^- -1}_{\bar{\l} \l}~,
\ee
which is proportional to the conjugate wave function (\ref{dualwave}).

\subsubsection*{Clebsh-Gordan Coefficients:}
A lot of important information, mathematical as well as  physical, is encoded
in the way the product of two wavefunctions decomposes as a linear sum
of wavefunctions.  On the mathematical side it contains all the
information we need about the tensor product of representations and
the recoupling coefficient involved.  
On the  physical side we can read out what the conservation rules are and how
two waves interact in our curved background.  We are interested in the
multiplicative properties of the normalized wave functions
\[
\phi^{p^+,p^-}_{\bar{\rho},\lambda}(g) = 
\sqrt{\frac{p^+}{\pi}}\, e^{-p^+ \l\bl} e^{-p^+ \r\brho} 
e^{ip^+(v+ia\bar{a}) +ip^-u} \, e^{2p^+ a\lambda e^{-iu}} 
e^{-2p^+ \bar{\rho} \bar{a}} \, e^{{2p^+}{\bar{\rho}}\lambda e^{-iu}}~.
\]
It is easy to see that    the product is given by
\bea
\phi^{p^+_1,p^-_1}_{\bar{\rho}_1,\lambda_1}(g)~
\phi^{p^+_2,p^-_2}_{\bar{\rho}_2,\lambda_2}(g) 
&=&\phi^{p^+_3,p^-_3}_{\bar{\rho}_3,\lambda_3}(g)~\sqrt{\frac{p_1^+p_2^+}{\pi \, p_3^+}} \, 
\exp [e^{-iu}({2 \, p^+_1}{\bar{\rho_1}\lambda_1} +{2 \, p^+_2}{\bar{\rho_2}\lambda_2}-{2 \, p^+_3} {\bar{\rho_3}\lambda_3})] \nn
&&\exp [ - \frac{p_1^+p_2^+}{p_3^+}(\l_1 -\l_2)(\bl_1 -\bl_2)] 
\exp [- \frac{p_1^+p_2^+}{p_3^+} (\rho_1 -\rho_2)(\brho_1 -\brho_2)]~, \nn
\eea
where we denote
$p^\pm_3 =p^\pm_1 +p^\pm_2 $, $p^+_3\l_3 = p^+_1\l_1 +p^+_2\l_2 $,
$p^+_3\rho_3=p^+_1\rho_1+p^+_2\rho_2$ as the momentum, position and radius
of the resulting wavefunction.  
The formulae for the addition  of ``position'' and ``radius'' follow also from the fact
that $p^+\lambda$ and $p^+\rho$  are
charges associated with  the left- and right-acting
isometries respectively. 
Intuitively  it is easier to explain as follows. In light-cone frame
$p^+$ is the ``effective mass'' for motion in the transverse  directions.
 Thus the above formula simply states that when
two particles coalesce into one, the resulting particle appears  at the centre of mass position.
 The same  intuitive explanation can  be offered to the ``radius'' addition rule if we go to a rotating
frame where the roles of radius and position are interchanged.

The term in the exponent can be evaluated to be
\be
\frac{p_1^+p_2^+}{p_3^+} \, ({\l_1} -{\l_2})
({\bar{\rho}_1}-{\bar{\rho}_2}) \, e^{-iu}~.
\ee
If we Taylor expand the exponential and  utilize the fact that
$e^{-iu} \phi^{p^+,p^-}_{\bar{\rho},\lambda} =
 \phi^{p^+,p^- - 1}_{\bar{\rho},\lambda}$,
 we get a simple result
\bea
\phi^{p^+_1,p^-_1}_{\bar{\rho}_1,\lambda_1}(g)~\phi^{p^+_2,p^-_2}_{\bar{\rho}_2,\lambda_2}(g) 
&=&\frac{1}{\sqrt{\pi}}
\exp(-\frac{p_1^+p_2^+}{p_3^+}|\rho_1 -\rho_2|^2) \exp(-\frac{p_1^+p_2^+}{p_3^+} |\l_1 -\l_2|^2)\nn
&&\sum_{n=0}^{\infty}
\frac{1}{n!}\left(\frac{p_1^+p_2^+}{p_3^+}\right)^{n+1/2}
({\l_1} -{\l_2})^n
({\bar{\rho}_1}-{\bar{\rho}_2})^n~
\phi^{p^+_3,p^-_3-n}_{\bar{\rho}_3,\lambda_3}(g)~.\nn
\eea
The wavefunction overlap is suppressed   exponentially  by the separation of
centres of two gaussians as one would expect. 
We  can also read out the tensorisation rules
\be
V^{p^+_1,p^-_1} \otimes V^{p^+_2,p^-_2} =
\sum_{n=0}^{+\infty} V^{p^+_1 +p^+_2,p^-_1+p^-_2 -n}~,
\ee
and the Clebsh-Gordan coefficients
\bea \label{waveCG}
C_{\l_1 \l_2 p^\pm_3}^{p^\pm_1 p^\pm_2 \l_3}
 &=& \sum_{n>0}  \delta( p_1^+ + p^+_2 -p^+_3) ~ \delta(p_1^- +p^-_2-p^-_3 -n)
 \delta^2(p_1^+\l_1+p^+_2\l_2 -p^+_3\l_3)  \nn
&& \times\frac{1}{\sqrt{n!}} \exp [- \frac{p_1^+p_2^+}{p_3^+} |\l_1 -\l_2|^2]
\left(\frac{p_1^+p_2^+}{p_3^+ }\right)^{\frac{n}{2} +\frac{1}{4}} ({\l_1} -{\l_2})^n ~.
\eea
Here we see that the $p^-$ momentum  is not conserved.
The change is in integral steps proportional to $H$, and
that the probability distribution is Poisson. Thus
for large values of separation/radii of the incoming particles
compared to string scale, we may wish to approximate
the Poisson distribution with
the normal distribution. The average shift in the value of $p^-$ is
\[
\bra \delta p^- \ket = {\frac{p_1^+p_2^+}{p_3^+} \, ({\l_1} -{\l_2})
({\bar{\rho}_1}-{\bar{\rho}_2})}
\]
and the width of the distribution is equal to the square root of $\bra \delta p^- \ket$ .

\section{Nappi-Witten Current Algebra and its Free Field Realization}
\label{sec:gammabeta}

Here we start solving our model in a covariant way.  To do that we propose a new free
field realization using the $\beta-\gamma$ system in Section~\ref{bg}.   
Next we construct the  currents and compute the corresponding algebra.  
In Section~\ref{ffcoordinate} we identify the original group coordinates 
 in terms of  the Wakimoto free fields and hence provide a geometric interpretation for latter.
Finally to pave the way for constructing the vertex operators we prove that
 the field identifications, done at the classical level, can be promoted to operator relations:
The OPEs of spacetime fields with the symmetry currents are given by the action
of the isometries.

\subsection{Nappi-Witten WZW model} \label{bg}

We now present  a new realization of the current algebra in terms of free fields,
comprised of a pair of null bosonic fields $u$, $\tilde{v}$ and the bosonic ghost
system $\beta$, $\gamma$.
A generic group element is written in terms of these fields
\be  \label{g-betagamma}
g = e^{\g J^-} e^{uJ+\tilde{v}T}  e^{\bgg J^+}~.
\ee
The inverse is very simple in this representation, and simplifies many computations:
\be
g^{-1} = e^{-\bgg \, J^+} \,  e^{-u \, J-\tilde{v}\, T} \,   e^{-\gamma \, J^-}~.
\ee
We can identify these variables with the geometric coordinates on the group
\be
\label{beta-gamma-a}
\gamma = \bar{a}~ ,\spa \bar\gamma = e^{-iu} \, a~,\spa  \tilde{v} = v + i \, a \, \bar{a}~.
\ee
It is straightforward to obtain the corresponding WZW action in terms of these variables:
\be \label{gammaaction}
S=\frac{1}{2\pi}  \,  \int d^2z
 \,(  \, \pd u  \, \bpd \tilde{v} +  2\, e^{iu}  \, \bpd\gamma  \, \pd \bar\gamma)~.
\ee
The metric and B-field are:
\bea \label{eq:gammabetaGandB}
ds^2 &=& 4 \, e^{iu} d\gamma \, d\bar\gamma + 2 \, du \,  d\tilde{v}   \\
B &=& 2e^{iu} d\gamma \wedge d\bar\gamma + du\wedge d\tilde{v}~.
\eea
We introduce two subsidiary fields $\beta$ and $\bar\beta$, which are the
canonical momenta associated with $\gamma$ and $\bar\gamma$.
In the usual first order formalism the action takes the form:
\be  \label{actionbetagamma}
S= \frac{1}{2\pi} \int \,  \left( \pd u \,  \bpd \tilde{v} +
\bar\beta \, \pd \bar\gamma +   \beta \, \bpd \gamma -
     \half \, e^{-iu}\,  \beta \,  \bar\beta \right)~.
\ee
The equations of motion enforce that $\beta = 2 \, e^{iu} \, \pd \bar\gamma$
and $\bar\b = 2 \, e^{iu} \,  \bpd \gamma$.

This form of the action is ultimately the basis of our solution to the Nappi-Witten model.
The first three terms make up the free action of the $\beta, \gamma$ ghost system ($c=2$)
 plus a pair of null light-cone fields ($c=2$). The last term,
usually called the screening charge in the literature, can be viewed
as an interaction term which however cannot contribute to any loop diagrams
on the worldsheet.

The symmetry currents $J_L(z)=g \, \pd g^{-1}$
and $J_R(\bz)= g^{-1} \,  \bpd g$ can also be conveniently rewritten
in terms of the first-order fields:
\be
\bary{rclrcl}\label{Jbetagamma}
  J^-_L(z)       &=& - \, \beta  &
  J^-_R(\bz)       &=&  2 \, (\bar\partial \bar\gamma + i \, \bar\partial u\, \bar\gamma)\\
  J^+_L(z)        &=&  -2 \, (\partial \gamma + i \, \partial u\, \gamma)&
 J^+_R(\bz)      &=&  \, \bar\beta  \\
  J_L(z)            &=& -(\partial \tilde{v} - \, i \,   \beta \,\gamma ) &
  J_R(\bz)        &=&  \bar\partial  \tilde{v} - i \, \bar\beta \,  \bar\gamma  \\
  T_L(z)           &=& - \partial u&
 T_R(\bz)   &=&  \bar\partial u~. \\ \eary
\ee
 From these expressions one may suspect  that $\beta$ and $\gamma$
are purely left-moving, whereas $\bar\beta$ and $\bar\gamma$ are purely right-moving. 
 This is indeed the case and will be clear from the field decompositions in~(\ref{coordef}).

Contracting  free fields according to
\bea \label{freefieldOPE}
u(z)       \,  \tilde{v}(w)              & \sim &  \ln (z-w)  \nn \cr
\beta(z) \,  \gamma(w) & \sim & \frac{1}{z-w}
\eea
the currents  satisfy the correct OPEs,
\bea    \label{OPEbetagamma}
          J (z) \, J^\pm (w) &\sim& \displaystyle \pm i \, \frac{ J^\pm(z)}{z - w} \nn
           J^+(z) \, J^-(w) &\sim& \displaystyle\frac{2}{(z-w)^2} +2 \, i \, \frac{T(z)}{z-w}\nn
           T(z) \, J (w)  &\sim& \displaystyle\frac{ 1}{(z - w)^2}~.
\eea
The Sugawara energy-momentum tensor also
admits a simple form:
\be
 {\cT}(z) =  \,  J(z) \, T(z) + \frac{J^+(z) \, J^-(z)}{2}=
 \beta(z) \, \partial\gamma(z) +\partial u \, \partial \tilde v.
\ee
 Readers who are interested in the
amplitudes may wish to skip ahead to Section~\ref{sec:correlation}.

\subsection{Free Field Representation of the Coordinates}
\label{ffcoordinate}

We have seen in the previous section how the Wakimoto free field
representation comes naturally from the Nappi-Witten action when
it is written in terms of the new coordinates and in first order
formalism.  However we have to pay a price for the algebraic  simplicity of this
representation,  we have lost the geometrical
interpretation  because the ghost $\beta$ is
not a coordinate field in spacetime.
In this section we will recover the geometrical interpretation of the free fields
 by expressing them in terms of the spacetime fields. Such geometrical representation
 for the $\gamma-\beta$ system has been proposed in the $SU(2)$ case 
 \cite{Falceto:1992bf,Bernard:1989iy}.

The Wakimoto free fields consist of the set of  holomorphic free
fields $u_L(z), \tilde{v}_L(z),$ $ \gamma_L(z), \beta_L(z)$ and
antiholomorphic free fields
$u_R(\bz),\tilde{v}_R(\bz),$ $\bar{\gamma}_R(\bz),
\bar{\beta}_R(\bz)$. We want to understand the relation of these
fields with the spacetime fields $a(z,\bz), \bar{a}(z,\bz),$
$u(z,\bz), v(z,\bz)$. This can be done if we look at the classical
solution of the Nappi-Witten model. It is well known
 that a general solution to a WZW model
is given by a product of left and right movers $g(z,\bz)=g_L(z) \,
g_R(\bz)$.
 If we introduce the left- and right-moving fields $u_{\LR},
\tilde{v}_{\LR}, \gamma_{\LR}, \bar{\gamma}_{\LR}$, parametrizing
group elements as in~(\ref{g-betagamma}),
 then this solution can be explicitly written as
\be \label{coordef}
\begin{array}{rcl}
a(z,\bz)     &=&\displaystyle{ e^{+{i}u_L(z)}\, [\bar{\gamma}_L(z) +e^{iu_R(\bz)} \bar{\gamma}_R(\bz)]}~,\\
\bar{a}(z,\bz)   &=&\displaystyle{ e^{-{i}u_L(z)}\, [e^{iu_L(z)}\gamma_L(z) + \gamma_R(\bz)] }~,\\
u(z,\bz)         &=& \displaystyle{ u_L(z) + u_R(\bz)}~,\\
v(z,\bz)+i\,a\,\bar{a}(z,\bz)  &=& \tilde{v}_L(z) + \tilde{v}_R(\bz) +2\,i \,\bar{\gamma}_L(z) \,\gamma_R(\bz)~.
\end{array}
\ee
It is then easy to check that this is a solution of the Nappi-Witten equations of motion.
The fields $\bar{\gamma}_R, \gamma_L$ do not possess any monodromy when going around the point $z=0$,
but the fields $\bar{\gamma}_L, \gamma_R$ do.

Because the current algebra~\eqr{Jbetagamma} has been realized by using only $\gamma_L, \beta_L$
and $ \bar{\gamma}_R, \bar\beta_R$, we would like to entirely purge the dynamical
field content of the theory of the fields $\bar{\gamma}_L, \gamma_R$.
%
This is possible to achieve by inverting the on shell relations satisfied by
the ghost fields $\beta_L(z), \bar{\beta}_R(\bz)$
\be
\begin{array}{rcl}
\beta_L(z)         &=& \displaystyle{2 \, e^{iu_L}}\, \partial_z {\bar{\gamma}}_L(z)~,\\
\bar{\beta}_R(\bz) &=& \displaystyle{2 \, e^{iu_R}}\, \bar{\partial}_{\bz}
{\gamma}_R(\bz)~.
\end{array}
\ee
We propose the following contour integrals in order to express the two remaining coordinates $\bar{\gamma}_L, \gamma_R$ in terms of the Wakimoto free fields
\be
\begin{array}{rcl}\label{screenint}
\bar{\gamma}_L(z) &=& \displaystyle{\frac{1}{\sin \pi p^+}
\oint_{-\pi}^{+\pi}\frac{d\sigma}{2 }~ e^{-iu_L(ze^{i\sigma})} ~ze^{i\sigma}~ \beta_L(ze^{i\sigma})}~,\\
{\gamma}_R(\bz) &=& \displaystyle{\frac{1}{\sin \pi p^+}
\oint_{-\pi}^{+\pi}\frac{d\sigma}{2 }~ e^{-iu_R(\bz e^{i\sigma})} ~ \bz e^{i\sigma}~ \bar\beta_R(\bz e^{i\sigma})}~,
\end{array}
\ee
where $p^+$ is the monodromy of the  chiral fields $u_L(z), u_R(\bz)$
when going around $0$: $ip^+= \frac{1}{2\pi} \oint dz \, \partial u(z)$.

\subsection{Quantum Free Field}
\label{qff}

Our goal in this section is to show that the previous
representation of the space time fields can be extended to the quantum level.
Namely, we want to show that the OPE of the currents with the spacetime fields 
can be interpreted in terms of the spacetime isometry.
This means that we now have to consider  the fields
 $u_{\LR}, \tilde{v}_{\LR}, \gamma_L,  \bar{\gamma}_R, \beta_L, \bar\beta_R$
  to be quantum operators.
The integrals (\ref{screenint}) above define
the composite operators $\bar\gamma_L, \gamma_R$. No operator ordering
is needed because $u(z)$ commutes with itself since it is a null
direction and it also commutes with $\beta$. We can now use the
expression~(\ref{coordef}) to define the coordinate fields $a,
\bar{a}, u, v$ as quantum fields. The coordinate fields
involve products of free fields at the same point, so we have to be
a little bit cautious in their definition. As before $u$ is
null and commutes with $\gamma_\LR, \bar\gamma_\LR$, so the only
problematic product is the one of $a $ with $\bar{a}$ since it
involves a product of $\beta$ with $\gamma$. We take care of this
issue in the usual way by normal ordering the $\beta$ and
$\gamma$. This means that the spacetime fields defined by
~(\ref{coordef},~\ref{screenint}) can be considered as quantum
operators. From the definition of the currents~(\ref{Jbetagamma}) in terms of
the Wakimoto free fields we can easily compute the OPEs of the
currents with the free fields.

Let $F(a,\bar{a}, u, v)(z,\bz)$ be a linear
functional of the fields (we take it to be linear in order to
avoid operator ordering problems).
For the left-moving fields
\be\label{consleftc}
\begin{array}{rcl}
T_L(z) ~  F(w,\bar{w}) &\sim & -\displaystyle{\frac{1}{z-w} ~ \dr_v  F(w,\bar{w})}~,\\
J_L (z) ~ F(w,\bar{w}) &\sim & -\displaystyle{\frac{1}{z-w} ~ (\dr_u + i(a\dr_a -\bar{a}\dr_{\bar{a}}))\, F(w,\bar{w})}~, \\
J_L^+(z) ~F(w,\bar{w}) &\sim & -\displaystyle{\frac{1}{z-w} ~ (\dr_{a}  + i  \ba \dr_v) \,F(w,\bar{w})}~, \\
J_L^-(z) ~F(w,\bar{w}) &\sim & -\displaystyle{\frac{1}{z-w} ~ (\dr_{\ba}- ia\dr_v)\, F(w,\bar{w})}~.
\end{array}
\ee
For the right-moving fields
\be\label{consrightc}
\begin{array}{rcl}
T_R(\bar{z}) ~  F(w,\bar{w}) &\sim  & \displaystyle{ \frac{1}{\bz-\bar{w}} ~ \dr_v F(w,\bar{w})}~,\\
J_R(\bar{z}) ~  F(w,\bar{w}) &\sim  & \displaystyle{ \frac{1}{\bz-\bar{w}} ~ \dr_u F(w,\bar{w})}~, \\
J^+_R(\bar{z}) ~F(w,\bar{w}) &\sim  & \displaystyle{ \frac{1}{\bz-\bar{w}} ~ e^{i\h u} (\dr_{a} - i \,\ba \, \dr_v ) F(w,\bar{w})}~, \\
J^-_R(\bar{z}) ~F(w,\bar{w}) &\sim  & \displaystyle{ \frac{1}{\bz-\bar{w}} ~ e^{-i\h u} (\dr_{\ba} +i \, a \,\dr_v) F(w,\bar{w})}~.
\end{array}
\ee
On the RHS of these equations we recognize the action of the
space isometries already discussed in (\ref{conscharge}).
This means that our construction of the spacetime coordinate fields
in terms of free fields is indeed the most natural one.
This construction is very different from the previous construction of
Kiritsis and Kounnas \cite{Kiritsis:jk}.
Moreover the geometrical interpretation of all the free fields, including
the $\beta$ field is now clear, which is usually one of the weak points of the
Wakimoto free field representation.

The proof of the OPEs (\ref{consleftc}, \ref{consrightc}) is obtained by a direct
and systematic computation.
Let us outline it for the right algebra.
 Using (\ref{freefieldOPE}) the nontrivial OPEs of $u, v, \bar\gamma$ with $J$ 
 are readily given by (we only list the right sector here):
\be
\begin{array}{rclrcl}
 J_R(\bar{z}) ~\bar{\gamma}_R(\bar{w}) & \sim & \displaystyle{-\frac{i \,\bar{\gamma}_R(\bar{w})}{\bz-\bar{w}}}~,& \spa
J_R(\bar{z})~u_R(\bar{w}) &\sim & \displaystyle{\frac{1}{\bz-\bar{w}}}~, \\
 J^-_R(\bar{z}) ~ \tilde{v}_R(\bar{w}) &\sim & \displaystyle{\frac{2\, i \,\bar{\gamma}_R(\bar{w})}{\bz-\bar{w}}}~,&\spa
T_R(\bar{z})~\tilde{v}_R(\bar{w})
&\sim & \displaystyle{\frac{1}{\bz-\bar{w}}}~,\\
J^+_R(\bar{z})~\bar{\gamma}_R(\bar{w}) &\sim & \displaystyle{\frac{1}{\bz-\bar{w}}}~.& & &
\end{array}
\ee
To compute the OPE of the currents with $\gamma_R$
one first computes the OPE of the currents with
$\bar{\partial}{\gamma}_R = 1/2\, e^{-iu_R}\, \bar{\beta}_R(\bz)$ and checks that
this object has a trivial OPE with all the currents except for $J_R^-$
\be
J_R^-(\bz) ~\frac{1}{2} e^{-iu_R}(\bar{w}) \bar{\beta}_R(\bar{w}) \sim
\bar{\partial}_{\bar{w}}\left(\frac{e^{-iu_R(\bar{w})}}{\bz -\bar{w}} \right)~.
\ee
Note that in the literature on Wakimoto free fields  this field is called the `screening current' and
${\gamma}_R$ the `screening charge'.
From the previous OPE one gets that the only nontrivial OPE of $J$ with $\gamma_R$ is given by
\be
J_R^-(\bz) ~\gamma_R(\bar{w}) \sim \frac{e^{-iu_R(\bar{w})}}{\bz -\bar{w}}~.
\ee

\section{Vertex Operators and Their Construction}
\label{sec:vertexop}

We have constructed a Wakimoto
representation of the current algebra in terms of free fields.  We now
want to use this free field representation  to define
the vertex operators and compute the correlation functions of the
theory.  This computation will give us a very simple and general
formula that we will consider more to be a conjecture than a full
proof.   we will in the next section
make use of the  free field representation to construct the vertex operators
as operators in the Hilbert space
and compute their matrix elements.  Before going on with the
computation let us give a description of the Nappi-Witten Hilbert
space and recall some general and well known facts about conformal
field theory and vertex operators.

\subsection{Nappi-Witten Hilbert Space and Vertex Operators}
\label{sec:hilbert}

One of the main properties of a two-dimensional conformal field theory is the fact that there is
a general  correspondence between states and operators.
This correspondence leads to the key notion of vertex operators as follows.
Given a state $|\phi\ket$ in the Hilbert space, there is a unique operator
$V_{|\phi\ket}(z,\bz)$   such that
\be
V_{|\phi\ket}(0)~ |0\ket =|\phi\ket,
\ee
with $|0\ket$ the $sl(2,C)$ invariant vacuum.
In order to describe the vertex operators we
first have to describe the Hilbert space of our theory.
We have seen that the Quantum mechanical Hilbert space
is given by the space of wave functions
\be
L^{2}(G) = \int_{0}^{+\infty}
 dp^{+} \int_{-\infty}^{+\infty} dp^- \left(V^{p^{+},\,p^-} \otimes \widetilde{ V}^{p^{+},\,p^-}
\oplus \widetilde{ V}^{p^{+},\,p^-} \otimes V^{p^{+},\,p^-}\right).
\ee
From the point of view of the CFT the Hilbert space is constructed out of
irreducible representations of the affine algebra.
It is well known that given a representation $V^{p^{+},\,p^-}$ of the Lie algebra we can
construct an highest weight representation of the current algebra
denoted ${\cal F}^{p^{+},\,p^-}$. These representations are generated by
action of the negative modes currents $J^a_{-n}$ on the vectors
$|\lambda\ket \in V^{p^{+},\,p^-}$ which are all annihilated by  positive modes currents
$J^a_{n} \, |\lambda\ket = 0 $, $n>0$.
Due to the timelike signature of  spacetime these representations are
not unitary. However in the Nappi-Witten model one can check that,
after imposition of the string mass shell conditions
$(L_0-1) \, |\psi\ket=0; L_n \, |\psi\ket=0, n>0$,
these representations are ghost free as long as $p^{+}<1$~\footnote{See \cite{Evans:1998wq}
for a clear  exposition of similar facts in $AdS_3$}.
This is already clear in the light-cone analysis of the model (Section~\ref{sec:lc}).
We will see that it is not possible to restrict to the sector with $p^+<1$ since long-string
states appear in the factorization of four-point amplitude (Section~\ref{sec:4point}).
Similar analysis had been done for strings in  $AdS_3$~\cite{maloog3}.
This shows that long-strings are part of the physical spectrum.
If $p^{+}=1$ the representation contains null states and if
$p^{+}>1$ the highest weight representation contains additional
negative-norm states\footnote{One can directly check
that the square norm of $J^-_{-1} \, |0\ket$ is given by $ 2\,(1-p^{+})$.}.
In order to construct ghost free representation of the current algebra
with $p^{+}>1$ we have to consider spectral flowed representation \cite{maloog1}.
First let us remark that the Nappi-Witten current algebra possesses an
automorphism {\cite{kirpio, D'Appollonio:2003dr} which is \\
\be
S_{w}(J^{\pm}_{n}) = J^{\pm}_{n \pm w}~, \spa
S_{w}(T_{n}) =T_{n}+ i \,w\, \delta_{n,0}~,\spa
S_{w}(J_{n}) =J_{n}~.
\ee
\\
Spectral flowed representations ${\cal F}^{p^{+},\,p^-}_w$ are defined as the highest weight
representations of the algebra obtained after action of this automorphism%
\footnote{A formal proof that these representations are ghost free is expected
to be very similar to the $\textrm{AdS}_3$ case but is not yet
available in the literature, see however \cite{Hikida:2002fp}.}.
The necessity of spectral flowed representation is clear from the
semi-classical analysis where they correspond to string winding around
the transverse plane\footnote{The automorphism comes
from the following action in the loop  group
$g(z,\bz) \rightarrow z^{iwJ}\,g\,{\bz}^{-iwJ}$}
\cite{maloog1, kirpio}.
We have also  seen in Section~\ref{sec:lc} that they
naturally appear in the spectrum of the light-cone Hamiltonian~\cite{Forgacs:1995tx}.
Overall, this leads to a Hilbert space
\be
{\cal H} = \sum_{w=0}^{\infty}
\int_{0}^{1} dp^{+} \int_0^{+\infty} dp^-
\left( {\cal F}^{p^{+},\,p^-}_w \otimes \widetilde{\cal F}^{p^{+},\,p^-}_w
\oplus \widetilde{\cal F}^{p^{+},\,p^-}_w \otimes {\cal F}^{p^{+},\,p^-}_w \right).
\ee
Given a state
$ |\bar{\l},\l\ket \in \widetilde{ V}^{p^{+},\,p^-} \otimes V^{p^{+},\,p^-}$
we can associate to it an highest weight state of
$\widetilde{\cal F}^{p^{+},\,p^-}_w \otimes {\cal F}^{p^{+},\,p^-}_w$
if we identify $w$ with the integer part of $p^+$.
A general state in the Hilbert space is obtained by repeated action
of the negative modes of the currents $J^i_{-n}$ on a highest weight state.

The vectors labelled  by nonzero integers $\omega$ are long-string states.
In these sectors the spectra are labelled by $0\leq p^+ <1$,  $p^-\in\mathbb{R}$.
The state corresponding to $p^+=0$ has a special status: it corresponds
to a long-string  vaccuum. 
This long-string state, as well as those related to $p^+=0$ state by spectrum 
flow~\cite{D'Appollonio:2003dr}, experiences  no potential in the transverse directions.
Therefore the representation they lie in are different from the massive long
string states with $p^+\neq 0$.
If we consider, as we do in this paper, wave packets (labelled by $p^+$)
localized in the transverse plane, 
these states appearing as one point in a continous spectrum  do not play a major role.
The representation for these states are  drastically
different from the massive $p^+$ representations, hence the construction of
vertex operators for these states is modified, and will the subject of a  future publication.
At present  we  concentrate only on massive $p^+\neq0$  states.

\subsection{Vertex Operators}
\label{sec:vertex}

In this section we want to construct the primary vertex operators
which are the vertex operators associated with the  highest weight states.
From the previous section discussion we now understand that there is a one to
one correspondence between the wave functions
$\phi_{\bar{\l},\l}^{p^{+},\,p^-}(g) \in \widetilde{V}^{p^{+},\,p^-} \otimes V^{p^{+},\,p^-}$,
and the primary vertex operators $V_{\bar{\l},\l}^{p^{+},\,p^-}(z,\bz)$
\footnote{Similarly one can associate a conjugate vertex operator
$\widetilde{V}_{\bar{\l},\l}^{p^{+},\,p^-}(z,\bz)$  to the conjugate wave functional
$\widetilde{\phi}_{\bar{\l},\l}^{p^{+},\,p^-}(g) \in { V}^{p^{+},\,p^-} \otimes \widetilde{V}^{p^{+},\,p^-}$.}.
We now want to construct this vertex operator restricting ourselves to the case
$0< p^+ <1$ of unflowed states.
We have seen in Section~\ref{qff} that the representation
of the coordinates in terms of free
fields introduced in Section~\ref{ffcoordinate} satisfy
the expected OPE with the currents.
It is interesting to reexpress the wave functional
$\phi_{\bar{\l},\l}^{p^{+},\,p^-}(g)$
in terms of the free fields, after substitution we obtain
\be
 V^L_{\bar{\l}}(z) ~ V_{\l}^R(\bz) ~ e^{-2p^+ S_{\bar{\l}\l}(z,\bz)}~,
\ee
where
\bea
V_{\bar{\l}}^L(z)= e^{ip^+ \vtil_L}\, e^{ip^- u_L}\, e^{-2p^+ \bar{\l}{\gamma}_L}~, \\
V_{\l}^R(\bz)= e^{ip^+ \vtil_R}\, e^{ip^- u_R}\, e^{2p^+ \l \bar{\gamma}_R}~,
\eea
and
\be
S_{\bar{\l}\l}(z,\bz) = (\gamma_R -\lambda e^{-iu_R})(\bar{\gamma}_L +\bar{\lambda} e^{-iu_L})~.
\ee
We want to promote this wave functional to a quantum vertex operator,
this is done by normal ordering the operators.
We denote as usual ${:}O{:}$ the normal ordering of an operator and define
\be
V_{\bar{\l},\l}^{p^{+},\,p^-}(z,\bz)= {:}V^L_{\bar{\l}}(z){:}~ {:}V_{\l}^R(\bz){:}~ {:}e^{-2p^+ S_{\bar{\l}\l}(z,\bz)(z,\bz)}{:}.
\ee
One can now check that {\it both}\, $V_{\bar{\l},\l}^{p^{+},\,p^-}(z,\bz)$ and
${:}V_{\bar{\l}}^L(z){:}$ satisfy the following OPE with the currents:
\bea\label{OPEl}
T_L (z) ~ V_{\bar{\lambda}}(w)  &\sim &   \frac{ -i\, p^+   }{z - w} \, V_{\bar{\lambda}}(w), \nn
J_L (z) ~ V_{\bar{\l}}(w)  &\sim &  \frac{   i \,( -p^- + \bar{\l} \partial_{\bar{\l}}) }{z - w} \, V_{\bar{\lambda}}(w),\nn
J^+_L(z) ~ V_{\bar{\l}}(w)&\sim &   \frac{-\partial_{\bar{\l}}}{z - w} \, V_{\bar{\lambda}}(w), \nn
J^-_L(z) ~ V_{\bar{\l}}(w) &\sim &   \frac{ 2 \, p^+\bar{\l}  }{z - w} \, V_{\bar{\lambda}}(w).
\eea
This OPE tells us that $V^{L}_{\bar\lambda}(z)$ satisfies the same OPE as the chiral vertex
operator associated with the state
\be
|{\bar{\l}}\ket= \exp( -\bar{\l} J^+) ~ |{0}\ket\in \tilV^{p^+,\,p^-}.
\ee
Similarly {\it both}\,  $V_{\bar{\l},\l}^{p^{+},\,p^-}(z,\bz)$ and
${:}{V}_{\l}^R(\bz){:}$  satisfy the same OPE with the right currents.
Therefore we should associate the state
\be
|\l\ket= \exp( -\l J^-)~|0\ket \in V^{p^+,\,p^-}
\ee
with the chiral operator, $V_\lambda^{R}(\bz)$.

Now both $ V_{\bar{\l},\l}^{p^{+},\,p^-}(z,\bz)$ and
${:}V_{\bar\l}^L(z){:}{:}{V}_{\l}^R(\bz){:}$
satisfy the same OPE's with the currents we thus expect them
to differ only by some screening operator
which commutes with the currents.
We have already seen that the current algebra admits a {\it unique} screening operator:
\be
S= \frac{1}{2}\, \int d^2\omega ~ \beta_L(\omega)\,\bar{\beta}_R(\bar\omega) \, e^{iu(\omega)}.
\ee
We have also seen that the Nappi-Witten action
written in first order formalism~(\ref{actionbetagamma})
is a free field theory plus an interaction term which is exactly $S$.
This suggests that the vertex operator can be written simply as
\be
V_{\bar{\l},\l}^{p^{+},\,p^-}(z,\bz) = {:}V_{\bl}^L(z)\, V_\l^R(\bz){:}~,
\ee
but at the expense of the necessity to insert the screening operator
into the expression for the correlation function.

\section{Correlation Functions}
\label{sec:correlation}

The conjecture above states that  in order to compute
the $N$-point correlation function  in the interacting theory
we just have to compute the free correlation function of the insertion of $N$
free vertex operators in the presence of the screening charge.
Similar approach for the case of $SL(2)$ has been used in
\cite{Becker:1993at,Andreev:1995bj,Petersen:1995rv,Teschner:1997ft,
Saraikin:1999cg,Giribet:2000fy,Hosomichi:2000bm}.

Thus the complete correlation function is given by
\be
\G_{\vec{\l},\vec{\bar{\lambda}},s}(\vec{z},\vec{\bar{z}})=
\sum_s \frac{1}{s!}
\left\bra\prod_{i=1}^n V_{\lambda_i}(z_i) ~ V_{\bar{\lambda}_i}(\bar{z}_i)
\left( \int d^2w ~\b(w)\,\bar{\b}(\bar{w})\, e^{-iu(w,\bar{w})} \right)^s \right\ket~~,
\ee
where $s$ is  a positive integer.
We have to integrate over $u_0,v_0,\gamma_0,\bar{\gamma}_0$ the zero modes of the fields
$u_L+u_R,v_L+v_R,\gamma_L, \bar{\gamma}_R$.%
\footnote{There is no integration over the zero modes of
$\beta_L$, $\bar{\beta}_R$ since they are identified with derivatives (momenta) of coordinate fields.}
The full partition function can then be split after integration over the
zero modes into a chiral and anti-chiral parts.
The invariant measure on the group is given by
\be
du_0 \, dv_0 \, d\gamma_0 \, d\bar{\gamma}_0 \, e^{iu_0}~.
\ee
The integration over the $u$ and $v$ zero modes
leads to conservation of momenta by producing delta functions
\be\label{pconsrule}
2 \pi \delta\left(\sum_{i=1}^n p^+_i \right) 2\pi \delta\left(\sum_{i=1}^n p^-_i+ 1 -s\right)~.
\ee
The integration over $\gamma_0, \bar{\gamma}_0$ also leads to momentum conservation
\be
\pi^2 \delta^{(2)}\left(\sum_{i=1}^n p^+_i \lambda_i \right)~.
\ee
The remaining term multiplying these delta functions is
\be
 \int  \prod_{k=1}^s d^2 w_k ~ |\G_{ \vec{\l}}(\vec{w},\vec{z})|^2~,
\ee
where $ \G_{\vec{\l}}(\vec{w},\vec{z})$ is a chiral conformal block
given by the product of a free field conformal block and a ghost conformal block.
 The free field block is given by
\be
\G^{uv}_{\vec{\l}}(\vec{w},\vec{z}) =
\left\bra\prod_{i=1}^n e^{ip^+_i \tilde{v}(z_i)+ i p^-_i u(z_i)}
\prod_{k=1}^s e^{-i u(w_k)}\right\ket~,
\ee
and can be easily computed by contracting the exponentials in all possible ways.
There are no $w_k,w_l$ contractions since $u$ is a null direction.
For each pair $z_i,z_j; i < j$ we get after contraction a factor $ (z_i -z_j)^{-p_i^+p_j^- -p_j^+p_i^-} $
since the free field OPE is given by $u(z) v(w) \sim \ln(z-w)$,
each pair $ z_i,w_k$ gives a factor $(w_k-z_i)^{p^+_i}$, overall
we get
 \be
\G^{uv}_{\vec{\l}}(\vec{w},\vec{z}) =
\prod_{i\neq j} (z_i -z_j)^{-p_i^+p_j^-} \prod_{i=1}^n \prod_{k=1}^s (w_k-z_i)^{p^+_i}~.
\ee
The ghost conformal block is given by
 \be
\G^{\b\g}_{\vec{\l}}(\vec{w},\vec{z}) =
\left\bra \prod_{i=1}^n e^{p_i^+ \l_i \g(z_i)}  \prod_{k=1}^s \b(w_k) \right\ket~.
\ee
It is computed by contracting each $\b(w)$ once with each factor  $\exp \l_i \g(z_i)$.
Each contraction gives a contribution $ \l (w-z)^{-1}$,
\be
\G^{\b\g}_{\vec{\l}}(\vec{w},\vec{z}) =
 \prod_{k=1}^s \left( \sum_{i=1}^n \frac{\l_i}{w_k-z_i}\right)~.
\ee
This block is homogeneous in $\l_i$ of degree $s$.

We can now put everything together and since there is no coupling between different
$w_i$ we get the simple expression
\bea    \label{corfunc}
\G_{\vec{\l},\vec{\bar{\lambda}},s}(\vec{z},\vec{\bar{z}})
=& 4\,\pi^4 ~\delta\left(\sum_i p^+_i\right)
\delta^{(2)}\left(\sum_{i=1}^n p^+_i \lambda_i \right)
\left(\prod_{i\neq j} |z_i -z_j|^{-2p_i^+p_j^-}\right) \nn
&\times \sum_{s\ge 0}^{} ~\delta\left(\sum_i p^-_i +1-s\right)~\frac{1}{s!}
\left(\cI_{\vec{\l},\vec{p}^+}(\vec{z})\right)^s~,
\eea
where
\be    \label{delta}
\cI_{\vec{\l},\vec{p}^+}(\vec{z}) = \int d^2w \left| \Delta_{\vec{\l},\vec{p}^+} ( w ,\vec{z})\right|^2 =
  \int d^2w \prod_{i=1}^n |w-z_i|^{2p^+_i}
\left| \sum_{i=1}^n \frac{p_i^+\l_i}{w-z_i} \right|^2~.
\ee
In this computation we have inserted $n$ vertex operators $V_{\bar{\l},\l}^{p^{+},\,p^-}$
associated with the wave function $\phi_{\bar{\l},\l}^{p^{+},\,p^-}$.
These vertex operators are well defined when $p^+$ is positive.
When $p^+$ is negative the appropriate vertex operator is the conjugate vertex operator
$\widetilde{V}_{\bar{\rho},\rho}^{p^{+},\,p^-}$ associated with $\widetilde{\phi}_{\bar{\l},\l}^{p^{+},\,p^-}$.
Due to the $p^+$ conservation rule not all $p^+$ momenta can be positive, at least one
should be negative. This means that one should at least insert one conjugate
vertex operator in  the correlation function.
Physically, particles with negative $p^+$ are necessary because they
correspond to outgoing states in the path integral prescription.
Fortunately, the conjugate vertex operator is related to the vertex operator we considered so far
by  an integral transform\footnote{${V}^{-p^+,-p^--1}$ appears instead of ${V}^{-p^+,-p^-}$
because the former is a solution to the wave equation with the same Casimir as ${V}^{p^+,p^-}$. 
See Appendix C for a rigorous treatment of the Integral Transform.} \eqr{duality}
\be \label{eq:dualV}
\widetilde{V}_{\bar{\rho},\rho}^{p^{+},\,p^-}
=
\frac{(p^+)^2}{\pi}\int d^2\l~e^{2p^+\l \rho}\, e^{2p^+\bar{\l} \bar{\rho}}~{V}^{-p^+,-(p^-+1)}_{\bar{\l}\l}~.
\ee
Using integration by parts in the right hand side of the OPEs
(\ref{OPEl}) one can easily show that
these operators satisfy with the currents the OPEs of a conjugate field.
This integral formula gives us the general rule needed
in order to insert a conjugate vertex operator in
the correlator: for each conjugate vertex operator one should
integrate over $\l$ a usual vertex operator
associated with momenta $-p^+,-(p^-+1)$.
In the case of one conjugate field we obtain
\bea   \label{corrVtV}
\left\bra
    \prod_{i=1}^{n-1}{V}^{p^+_i,\,p^-_i}_{\bar{\l}_i\l_i}(z_i)
    \widetilde{V}^{p^+_n,\,p^-_n}_{\bar{\rho}_n\rho_n}(z_n)
\right\ket
&=& \frac{1}{\pi}\delta\left(\sum_i p^+_i\right) |e^{\rho_n(\sum_{i=1}^{n-1} 2 p_i^+ \l_i)}|^2
    \prod_{i\neq j} |z_i -z_j|^{-2p_i^+p_j^-}   \nn
  &\cdot&\sum_{s\ge 0} ~\delta(\sum_{i=1}^{n-1} p^-_i -p_n^- - s)~\frac{1}{s!}
      \left(\cI_{\vec{\l},\vec{p}^+}(\vec{z})\right)^s~.
\eea
with $p_n^+\l_n =\sum_{i=1}^{n-1}p_i^+\l_i$ in $\cI_{\vec{\l},\vec{p}^+}$.

\subsection{Three-Point Function}

We can now specialize our general formula to the case
where there is only three points $z_1,z_2,z_3$.
In this case one can check that
\bea
\cI_{\vec{\l},\vec{p}^+}(\vec{z})&=&
|z_1-z_2|^{-2p_3^+} |z_1-z_3|^{-2p_2^+} |z_2-z_3|^{-2p_1^+} \nn
&\times& |\l_1 -\l_2|^2 (p_1^+)^2 \int d^2\tilde{w} |\tilde{w}|^{2(p_1-1)} |\tilde{w}-1|^{2p_2^+}~.
\eea
One first uses the identity $\sum_i p_i^+ \l_i=0$ to eliminate $\lambda_3$, then
makes the change of variable
\be
w \rightarrow \tilde{w} = \frac{(z_2-z_3)(w-z_1)}{(z_2-z_1)(w-z_3)},
\ee
and finally uses the fact that
\be
w^{p_1^+}(w-1)^{p_2^+}\left(\frac{p_1^+ \l_2}{w} +\frac{p_2^+\l_2}{w-1}\right)
\ee
is a total derivative which we can subtract from the integrand.
The integral can be explicitly evaluated
\be    \label{int}
\int d^2w ~|w|^{2(p^+_1-1)}\,|w-1|^{2(p^+_2-1)} = \pi \,\frac{\U(p_1)\U(p_2)}{\U(p_1 +p_2 )}~,
\ee
where $\U(p)\equiv \frac{\G(p)}{\G(1-p)}$. This function satisfies
$ \U(p)\U(1-p)=1$, $ \U(p+1) = -p^2 \U(p)$.
Therefore, using the value of the integral and the conservation of the light cone momenta
we can write this integral in a symmetric form
\bea
\cI_{\vec{\l},\vec{p}^+}(\vec{z}) &=&\pi |z_1-z_2|^{-2p_3^+} |z_1-z_3|^{-2p_2^+} |z_2-z_3|^{-2p_1^+} \nn
        &\times& \left|\frac{\l_1 -\l_2}{p^+_3}\right|^2 {\U(p_1^++1)\U(p_2^++1)\U(p_3^+ +1)}.
\eea

We can now easily read out from \eqr{corrVtV}  the  three-point function
\bea  \label{3point}
&&\left\bra {V}^{p^+_1,\,p^-_1}_{\bar{\l}_1\l_1}(z_1){V}^{p^+_2,\,p^-_2}_{\bar{\l}_2\l_2}(z_2)
 \widetilde{V}^{p^+_3,\,p^-_3}_{\bar{\rho}_3\rho_3}(z_3)\right\ket  \nn
&=&\frac{1}{\pi}\displaystyle{|z_1-z_2|^{-2\Delta_1 -2\Delta_2 +2\Delta_3}
                       |z_1-z_3|^{-2\Delta_1 -2\Delta_3 +2\Delta_2}
                       |z_2-z_3|^{-2\Delta_2 -2\Delta_3 +2\Delta_1}} \nn
&& \delta(p_1^++p_2^+-p^+_3) \,\, \displaystyle{|e^{2 \rho_3(p_1^+ \l_1 + p_2^+ \l_2)}|^2} \nn
&&\displaystyle{\sum_{s\ge0}\delta(p_1^-+p_2^- -p_3^- -s)} 
\frac{|{\l_1 -\l_2}|^{2s}}{\Gamma(s+1)}
 \displaystyle{\left(-\pi \frac{\U(p_1^++1)\U(p_2^++1)}{\U(p_3^++1)}\right)^s} 
\eea
where $\Delta_i =-p^+_i(p^-_i+1/2)$ and $s=p_1^-+p_2^--p_3^-$.

The $\delta$-functions on the right-hand side enforce the conservation rules.
The $\rho$, $\lambda$ dependence is as expected from the Clebsh-Gordan coefficients;
and  the global conformal invariance dictates the $z$ dependence.
 The only new   piece of information is momentum-dependent coupling constants:
\be \label{3coupling}
C_{p_1^-,p_2^-,p_3^-}^{p_1^+,p_2^+,p_3^+}= \displaystyle \frac{1}{s!}
\left(-\pi \frac{\U(p_1^++1)\U(p_2^++1)}{\U(p_3^++1)}\right)^{s}.
\ee
This result agrees with \cite{D'Appollonio:2003dr}.

We should  stress that the construction of the vertex
operator and the computation of the correlation function is a
priori valid only when $0<p_i^+<1$. However
the three point function is a holomorphic function with poles when
$p_3^+$ is a nonzero integer \footnote{$\U(1+p) $ admits a pole
when $p =-n$, $n\geq 1$, and a zero when $p =n$, $n\geq 0$.}. We are
going to see in the next section that for $p_i^+>1$  the three point
function gives the correct amplitude associated  with  long string states.

\subsubsection*{Integral Transform:}
We pause here to give some justification for the
prescription~\eqr{eq:dualV} which tells us that the correlation
functions with insertion of more than one conjugate vertex
operator is related by an integral transform to the correlation
function with only one conjugate vertex operator.
We expect the conjugate vertex operator $\widetilde{V}$ to be
related to $V^\dagger$, the hermitian conjugate of $V$.
Such an expectation is proved to be true due to the following identity for the
three points function
\bea  \label{tildedag}
&&\left\bra
   {V}^{p^+_1,\,p^-_1}_{\bar{\l}_1\l_1}(z_1){V}^{p^+_2,\,p^-_2}_{\bar{\l}_2\l_2}(z_2)
   \frac{\pi^{2p_3^-}\widetilde{V}^{p^+_3,\,p^-_3}_{\bar{\l}_3\l_3}(z_3)}{2^{2p_3^-}\U(p_3^++1)}
 \right\ket   \nn 
 &&\nn
 &=& -\pi^2 \left\bra
 \frac{\pi^{2p_1^-}\widetilde{V}^{p^+_1,\,p^-_1}_{\bar{\l}_1\l_1}(z_1)}{2^{2p_1^-}\U(p_1^++1)}
 \frac{\pi^{2p_2^-}\widetilde{V}^{p^+_2,\,p^-_2}_{\bar{\l}_2\l_2}(z_2)}{2^{2p_2^-}\U(p_2^++1)}
      {V}^{p^+_3,\,p^-_3}_{\bar{\l}_3\l_3}(z_3)
 \right\ket,
\eea
which shows that
\be
\widetilde{V}^{p^+,p^-} 
= -\frac{1}{\pi^2} \left(\frac{2}{\pi}\right)^{2p_1^-}\U(p_1^++1)({V}^{p^+,p^-})^\dagger.
\ee
We first express $ \bra \tilV_1 \tilV_2 V_3\ket$,  the right hand side of~\eqr{tildedag},
as an integral transform:
\be
\cJ  \equiv \frac{(p^+_2)^2}{\pi}\int {d^2\rho_2} ~e^{2p^+_2\l_2 \rho_2}\,
            e^{2p^+_2\bar{\l}_2 \bar{\rho}_2}~
       \left\bra \widetilde{V}^{p^+_1,\,p^-_1}_{\bar{\l}_1\l_1}(0)
              {V}^{-p^+_2,\,-p^-_2-1}_{\bar{\rho}_2\rho_2}(1)
              {V}^{p^+_3,\,p^-_3}_{\bar{\l}_3\l_3}(\infty)
       \right\ket~.
\ee
The $p^+$ conservation rule reads $ p_1^+ +p_2^+ -p_3^+=0$.
After the  change of variables:
$\eta = 2p_2^+(\l_3-\rho_2)(\l_2-\l_1)$ the integral transform, $\cJ$, reads
\bea
\cJ&=& \frac{(p^+_2)^2}{\pi}\displaystyle{|e^{2\l_3(2p_1^+ \l_1 + 2p_2^+ \l_2)}||\l_2-\l_1|^{2s}(2p_2^+)^{2s}}  \nn
    &&\frac{1}{\Gamma(-s)}\left(-\pi \frac{\U(p_3^++1)\U(-p_2^++1)}{\U(p_1^++1)}\right)^{-s-1}
       \int \frac{d\eta d\bar{\eta}}{\pi} |e^{-\eta} \eta^{-s-1}|^2
\eea
where we have defined $s=p_1^-+p_2^--p_3^-$.
The $\eta$ integral is evaluated to be\footnote{See Appendix~\ref{Itransf} for details.}
$(-1)^{s+1} \Gamma(-s)/\Gamma(s+1)$.
Finally we use the identity $ (p_2^+)^2/\U( -p_2^++1) = -\U(p_2^+ +1)$ to express
\bea
\cJ &=& \displaystyle -\frac{1}{\pi^2} \frac{\left(\frac{2}{\pi}\right)^{2p_1^-}\U(p_1^++1)
\left(\frac{2}{\pi}\right)^{2p_2^-}\U(p_2^++1)}{\left(\frac{2}{\pi}\right)^{2p_3^-}\U(p_3^++1)} \nn
&& \displaystyle \left|e^{2\l_3(2p_1^+ \l_1 + 2p_2^+ \l_2)}\right||\l_2-\l_1|^{2s}
\frac{1}{\Gamma(s+1)}\left(-\pi \frac{\U(p_1^++1)\U(p_2^++1)}{\U(p_3^++1)}\right)^{s}.
\eea
We recognize in the first line the normalisation factor which enters (\ref{tildedag})
and in the second the correlator $ \bra V_1 V_2 \widetilde{V}_3\ket$.  This proves our claim.
\\

{{\noindent\bf{Normalization:}}}
Given the symmetry in lightcone time reversal, {\it{i.e.}}
 $u \rightarrow -u$,
it is natural to look for a normalized  vertex operator, 
$\hV$, and its conjugate, $\hV^\dagger$, such that the three-point function
 are $CPT$ invariant:
\be 
\displaystyle\bra \hV_1 \hV_2 {\hV}_3^\dagger\ket= 
\displaystyle\bra {\hV}_1^\dagger {\hV}_2^\dagger \hV_3 \ket
\ee
and the two-point amplitude satisfies
\be
\bra \hV^{p_,^+ p^-} (\hV^{-p_,^+ -p^-})^\dagger \ket= 1~.
\ee
In light of (\ref{tildedag}) we realize that
\bea \label{normV}
\hV^{p^+,\,p^-}_{\bar{\lambda} \lambda} 
&=&\left(\frac{2}{\pi}\right)^{p^-} (-\U(p^+ +1))^{\half}e^{-2p^+\lambda \bar{\lambda}} 
        V^{p^+,\,p^-}_{\bar{\lambda} \lambda}, \\
({\hV}^{p^+,\,p^-}_{\bar{\lambda} \lambda})^\dagger 
 &=&\pi \left(\frac{\pi}{2}\right)^{p^-}
     (-\U(p^+ +1))^{-\half} e^{-2p^+\lambda \bar{\lambda}}\widetilde{V}^{p^+,\,p^-}_{{\lambda} \bar{\lambda}}~.
\eea
\\

{\noindent\bf{Two-Point Function:}}
The unit field is obtained from $V^{p^+,\,p^-}$ in the limit where $p^+=0$, $p^-=0$~\footnote{
Note $\hV$ blows up in this limit.}. 
We can therefore obtain the two point function by considering this limit in the three point function.
Starting from the expression  \eqr{3point} the limit is easily taken.
Since $\U(p+1) \rightarrow -p$ as $p$ approaches zero, only the term $s=0$ survives.
For the other factors entering the three-point function the limit can be trivially
taken.  One finds
\be  \label{2point}
\left\bra {V}^{p^+_1,\,p^-_1}_{\bar{\l}_1\l_1}(z_1)\widetilde{V}^{p^+_2,\,p^-_2}_{\bar{\rho}_2\rho_2}(z_2)\right\ket  \nn
=\frac{1}{\pi}\displaystyle{|z_1-z_2|^{-4\Delta}
\delta(p_1^+-p_2^+)\delta(p_1^--p_2^-)}
   \displaystyle{|e^{2 p_1^+ \l_1  \rho_2 }|^2}
\ee
where $\Delta= -p_1^+(p_1^- + \half)$ as before.
Using the normalized vertex operators, $\hV$, we have indeed
\be
\left\bra {\hV}^{p^+_1,\,p^-_1}_{\bar{\l}_1\l_1}(0)
({\hV}^{p^+_2,\,p^-_2}_{\bar{\rho}_2\rho_2})^\dagger(1)\right\ket 
=\frac{1}{\pi} \delta(p_1^+-p_2^+) \delta(p_1^--p_2^-)
   \displaystyle {e^{-2 p_1^+ |\l_1 - \rho_2|^2 }}~.
\ee
This completes the discussion on normalization of the vertex operators.

\subsection{Four-Point Function: 3--1}
\label{sec:4point}

The four-point function describing $3 \rightarrow 1$ scattering,
\be
\left\bra
{V}^{p^+_1,\,p^-_1}_{\bar{\l}_1\l_1}(z_1) {V}^{p^+_2,\,p^-_2}_{\bar{\l}_2\l_2}(z_2)
{V}^{p^+_3,\,p^-_3}_{\bar{\l}_3\l_3}(z_3)
\widetilde{V}^{p^+_4,\,p^-_4}_{\bar{\rho}_4\rho_4}(z_4)
\right\ket
\ee
is expressed in terms of the integral, $\cI_{\vec{\l},\vec{p^+}}(\vec{z})$,
 with  $p^+_4 = p_1^+ +p_2^++p_3^+$ and $ p^+_4\l_4 = p_1^+\l_1 +p_2^+\l_2 + p_3^+\l_3$.
Using the properties  of $\cI$ under the projective transformations of $z$  we get
\be
\cI_{\vec{\l},\vec{p}^+}(\vec{z}) =
 \prod_{i<j} |z_i-z_j|^{p_i^++p_j^+}
|z|^{-(p^+_1 +p_2^+)}|1-z|^{-(p^+_2 +p_3^+)}
I_{\vec{\l}}(z)
\ee
where
$ z $ being the cross ratio
$\displaystyle{z=\frac{(z_2-z_1)(z_3-z_4)}{(z_2-z_4)(z_3-z_1)}}$
and
\bea
I_{\vec\l}(z)&\equiv& \int d^2w |\Delta_{{\vec\l},z}({w})|^2  \nn
    &=& \int d^2w \left| w^{p_1^+} (w-z)^{p^+_2} (w-1)^{p^+_3}
      (\frac{p_1^+\l_1}{w} +\frac{p_2^+\l_2}{w-z} +\frac{p_3^+\l_3}{w-1})\right|^2~.
\eea
According to the general rules for chiral splitting of integrals presented in Appendix~\ref{app:split},
this integral can be written as
\bea
I_{\vec\l}(z)= \frac{\sin\pi p_1^+ \sin \pi p_2^+ }{\sin \pi(p_1+p_2)}
         \left|\int_0^z |\Delta_{{\vec\l},z}({x})| dx\right|^2
         + \frac{\sin\pi p_3^+ \sin{\pi p_4^+}}{\sin{\pi(p_1+p_2)}}
           \left|\int_1^\infty |\Delta_{{\vec\l},z}({x})| dx\right|^2  \nn
\eea
The line integrals can in turn be expressed in terms of hypergeometric functions
\bea
 \int_0^z |\Delta_{{\vec\l},z}({x})|
&=& \frac{\Gamma(p_1^++1)\Gamma(p_2^++1)}{\Gamma({p_1^++p_2^++1})} f_{\vec{\l}}(z)~,\\ \nn
 \int_1^\infty |\Delta_{{\vec\l},z}({x})|
&=& \frac{\Gamma(-p_4^+)\Gamma(p_3^++1)}{\Gamma({-p_1^+-p_2^+})}g_{\vec\l};
\eea
with $ f_{\vec{\l}}(z)=\sum_{i=1}^3 \l_i f_{i}(z)$,
$g_{\vec\l}=\sum_{i=1}^3 \l_i g_{i}(z)$ where
\bea
f_{1}(z)&\equiv &  {z^{p_1^++p_2^+}} F(-p_3^+,p_1^+,p_1^++p_2^++1;z),\nn
f_{2}(z)&\equiv &   - {z^{p_1^++p_2^+}} F(-p_3^+,p_1^++1,p_1^++p_2^++1;z),\nn
f_{3}(z)&\equiv & \frac{z^{p_1^++p_2^++1} p_3^+}{p_1^+ +p_2^++1}
     F(1-p_3^+,p_1^++1,p_1^++p_2^++2;z),
\eea
\bea
g_{1}(z)&\equiv&\frac{-p_1^+}{p_1^+ +p_2^+}  F(-p_2^+,-p_4^+,-p_1^+- p_2^+ +1;z), \nn
g_{2}(z)&\equiv&\frac{-p_2^+}{p_1^+ +p_2^+}  F(1-p_2^+,-p_4^+,-p_1^+- p_2^+ +1;z),\nn
g_{3}(z)&\equiv&  F(-p_2^+,-p_4^+,-p_1^+- p_2^+ ;z)~.
\eea
Note that these functions satisfy the Gauss recursions identities
\be
\sum_{i=1}^3  g_{i}(z)=\sum_{i=1}^3  f_{i}(z)=0.
\ee
Overall this gives
\bea
I_{\vec\l}(z)= -\pi\left( \frac{\U(p_1^++1)\U(p_2^++1)}{\U(p_1^++p_2^+ +1)} |f_{\vec{\l},\vec{p}}(z)|^2
+ \frac{\U(p_1^+ +p_2^++1)\U(p_3^++1)}{\U(p_4^++1)} |g_{{\vec\l},\vec{p}}(z)|^2\right)~. \nn
\eea

The total amplitude is then given by
\be
\left\bra
 {V}^{p^+_1,\,p^-_1}_{\bar{\l}_1\l_1}(0){V}^{p^+_2,\,p^-_2}_{\bar{\l}_2\l_2}(z)
 {V}^{p^+_3,\,p^-_3}_{\bar{\l}_3\l_3}(1)
 \widetilde{V}^{p^+_4,\,p^-_4}_{\bar{\rho}_4\rho_4}(\infty)
\right\ket
 =  e^{\rho_4(2p^+_1\l_1 +2p^+_2\l_2 + 2p^+_3\l_3)} \frac{(I_{\vec\l}(z))^s}{\Gamma(s+1)}
\ee
with $s= p^-_1 +p_2^-+p_3^- - p_4^-$ being an integer~\footnote{Due to the presence of
the gamma function in the denominator the amplitude is zero if $s$ is not a positive integer}
and  $p_1^+ +p_2^+ +p_3^+ =p_4^+$.
The four point function can now be expressed as a linear combination of conformal blocks
\be
\frac{ (I_{\vec\l}(z))^s}{\Gamma(s+1)} =
\sum_{\eta=0}^{s} C_{p_1^-,p_2^-,p_1^- + p_2^- -\eta}^{p_1^+,p_2^+,p_1^+ + p_2^+}\,
C_{p_1^- + p_2^- -\eta,p_3^-,p_4^-}^{p_1^+ + p_2^+ ,p_3^+,p_4^+}\,
|f_{{\vec\l},\vec{p}}(z)|^{2s}| \, g_{{\vec\l},\vec{p}}(z)|^{2(s-\eta)}
\ee
where  $C_{p_1,p_2,p_3}$ is the three-particle coupling constant~\eqr{3coupling}.

The factorization of the four point function shows that if we start with
short string vertex operators having $0<p_i^+<1$ there will be no intermediate states
corresponding to long strings since $ p_1^+ +p_2^++p_3^{+}=p_{4}^{+}<1$.
In order to have long string states propagating in the intermediate
channel we either have to insert external long string states $p_i^+>1$
{\it or} insert external state with, for instance $0>p_2^+>-1$, by our
integral transform rule this correspond to inserting one more conjugate
short string vertex operator.


\subsection{Four-Point Function: 2--2}
We now want to construct the amplitude with two conjugate fields, such an amplitude is
related by an integral transform to the previous amplitude
\bea
&&\left\bra
V^{p^+_1,\,p^-_1}_{\bar\l_1\l_1}(0) \widetilde{V}^{p^+_2,\,p^-_2}_{\bar\rho_2\rho_2}(z)
 V^{p^+_3,\,p^-_3}_{\bar{\l}_3\l_3}(1) \widetilde{V}^{p^+_4,\,p^-_4}_{\bar\rho_4\rho_4}(\infty)
\right\ket                          \nn
&=&e^{\rho_4(2p^+_1\l_1 + 2p^+_3\l_3)}\int d^2\l_2 \left| e^{2p^+_2\l_2 (\rho_2 -\rho_4)}\right|^2
 \frac{(I_{\vec\l}(z))^s}{\Gamma(s+1)}.
\eea
where $I_{\vec\l}(z)$ is defined as in the previous section except that we have to change
$p_2^+$ into $-p_2^+$ in  all the expressions and  $s = p_1^--p_2^-+p_3^--p_4^--1$.
We have:
\be
I_{\vec\l}=C_{12}|f_{\vec\l}|^2 +C_{34}|g_{\vec\l}|^2
\ee
where
\bea
C_{12} &=& \displaystyle \frac{\U(p_1^++1)\U(-p_2^++1)}{\U(p_1^+-p_2^+ +1)},\\
C_{34}&=&\displaystyle \frac{\U(-p_4^+)\U(p_3^++1)}{\U(p_2^+ -p_1^+)}.
\eea
For convenience we introduce the shorthand notations:
\bea
A &=& C_{12}|f_2|^2 + C_{34}|g_2|^2, \\
B &=& (\l_1 -\l_3)(C_{12} f_2 f_1 + C_{34}g_2 g_1)-\l_3 A,\\
AC-B\bar{B} &=& C_{12}C_{34}|\l_1-\l_3|^2|f_2g_1-g_2f_1|^2.
\eea
to write $I_{\vec\l}= A \l_2 \bar{\l}_2 +\bar{B}\l_2+B\bar{\l}_2 +C$.

In Appendix \ref{Itransf} we have computed such an integral transform.
Using the result of  (\ref{mainres})  we obtain the following expression
for the four-point function with 2 conjugate fields
\be\label{22}
e^{\rho_4(2p^+_1\l_1  + 2p^+_3\l_3)} \left|e^{-2p_2^+(\rho_2-\rho_4) \frac{B}{A}}\right|^2
\left(  \frac{D}{-4p_2^+|\rho_2 -\rho_4|^2}\right)^{s+1}
\frac{1}{A}  I_{-s-1}(\frac{D}{A}),
\ee
where we have denoted
$D\equiv 2p_2^+|\rho_2-\rho_4|\sqrt{B\bar{B}-AC}$ and $I_{-s-1}$ the modified Bessel function
of the first kind.
 We can evaluate this quantity to be
\be
D = \sqrt{-C_{12}C_{34}}~\left|2p_2^+(\rho_2-\rho_4)(\l_1-\l_3) z^{p_1^+-p_2^+}(1-z)^{p_3^+-p_2^+}\right|~.
\ee
In order to show this we first check that
\bea
   \frac{p_2^{+}¥}{z(1-z)} f_1(z) = a(z)^{-1} \partial_z( a(z)f_2(z)), \nn
   \frac{p_2^{+}¥}{z(1-z)} g_1(z) = a(z)^{-1} \partial_z( a(z)g_2(z))
\eea
where $a(z) = z^{-p_1^+}(1-z)^{p_1^+-p_4^+}$. This shows that
$AC-B\bar{B}$ is just a Wronskien which can be evaluated to be
\be
AC-B\bar{B} =C_{12}C_{34}|(\l_1-\l_3) z^{p_1^+
-p_2^+}(1-z)^{p_3^+-p_2^+}|^2.
\ee
The expression (\ref{22}) of the correlation function is similar to
the one obtained in \cite{D'Appollonio:2003dr}: Up to a $p^+$ dependent proportionality
factor it has the same $z$ dependence.
The factorisation of this amplitude performed in \cite{D'Appollonio:2003dr} can thus be applied
to our case and indeed shows that long strings propagate as intermediate states.

\section{Ward Identities and Flat Space Limit}
\label{sec:ward}

In this section we shall check that the N-point function we proposed satisfies
all the required Ward identities.  Our amplitudes also coincide with
 those in the flat space  upon sending $H$ to zero.
Let us first denote the chiral part of the correlation function as
\be
G_{\lambda,C}(\vec{z})\equiv
\left(\prod_{i\neq j} (z_i -z_j)^{-p_i^+p_j^- }\right)
\left( \oint_C dw \Delta_{\vec{\l},\vec{p}^+}({w},\vec{z})\right)^{s}~,
\ee
where C is a contour of integration.

\subsection{Conformal Ward Identities}

We can easily compute the transformation rule of $G_{\lambda,C}(\vec{z})$ and
our n-point function (\ref{corfunc}) under the projective transformations
\be\label{projtrans}
z\rightarrow \tilde{z} =\frac{(az+b)}{(cz+d)}.
\ee
Under such transformation $w-z_i$ becomes
$(w-z_i)/(cw+d)(cz_i +d)$ and $dw$ becomes $dw/(cw+d)^2$.
Therefore the factor $\prod_{i\neq j} (z_i -z_j)^{-p_i^+p_j^- -p_j^+p_i^-}$
is transformed as
\be
\prod_i(cz_i+d)^{p_i^+ (\sum_{j\neq i} p^-_j) + (\sum_{j\neq i} p^+_j)p_i^-}
  \prod_{i\neq j} (z_i -z_j)^{-p_i^+p_j^-}
  \ee
  using the conservation rules for the momenta \eqr{pconsrule} this gives
\be
\prod_i(cz_i+d)^{-2 p_i^+ p^-_i + p_i^+(s -1 )}
  \prod_{i\neq j} (z_i -z_j)^{-p_i^+p_j^-}.
  \ee
The integral
$\oint_C dw \Delta_{\vec{\l},\vec{p}^+}({w},\vec{z})$ under a projective transformation becomes
\be
\prod_i(cz_i+d)^{-p^+_i}
\oint_C dw \prod_{i=1}^n (w-z_i)^{p^+_i}
\left( \sum_{i=1}^n \frac{p_i^+\l_i (cz_i+d)}{(w-z_i)(cw+d)}\right).
\ee
One uses the identity
\be
\frac{ (cz_i+d)}{(w-z_i)(cw+d)} =
\frac{1}{w-z_i} -\frac{c}{cw+d}
\ee
and the conservation rules $\sum_i p_i^+ \l_i =0$ to see that the integrand is exactly
$dw \Delta_{\vec{\l},\vec{p}^+}({w},\vec{z})$.
Overall one gets
\be
G_{\lambda,C}(\frac{a\vec{z}+b}{c\vec{z} +d})=
\prod_{i=1}^n(cz_i+d)^{-2 p_i^+ p^-_i - p_i^+}
G_{\lambda,C}(\vec{z}).
\ee
which shows that the conformal weight of the vertex operator $V_\lambda^{p^\pm}$ is given by
\be
-p^+(p^-+1/2)
\ee
as expected.

\subsection{Algebraic Ward Identities}

For each generator of the Lie algebra $J^a=T,J,J^+,J^-$, we expect a conservation rule
$(\sum_i J^a_i) \G_{\vec{\l},\vec{\bar{\lambda}},s} =0$.
The $T$ conservation rule is the conservation of light-cone momenta
$\sum_i p^+_i =0$.
The $J^-$ conservation rule is equivalent to $\sum_i p^+_i \lambda_i=0$.
The $J^+$ conservation rule amounts to invariance under translation in $\lambda$,
$\G_{\vec{\l},\vec{\bar{\lambda}},s} =\G_{\vec{\l} +a,\vec{\bar{\lambda}},s}$.
This is clear first since $\sum_i p_i^+ a =0$ and also since
\be
\prod_{i=1}^n (w-z_i)^{p^+_i}
\left( \sum_{i=1}^n \frac{p_i^+ a }{(w-z_i)}\right)
= a \frac{\partial}{\partial w}\left(\prod_{i=1}^n (w-z_i)^{p^+_i}\right).
\ee
The $J$ conservation rule comes from the fact that $\G$ is homogeneous of degree
$s= \sum_ip_i^-$,
\be
\G_{a \vec{\l},\vec{\bar{\lambda}},s} =a^{\sum_ip_i^-}\G_{\vec{\l},\vec{\bar{\lambda}},s}.
\ee
This is clear since $\delta(\sum_ip_i^+\l_i) $ is homogeneous of degree $-1$ and
$ \Delta_{\vec{\l},\vec{p}^+}({w},\vec{z})$ is  homogenous of degree $+1$ in $\lambda$.

\subsection{Knizhnik-Zamolodchikov Equation}

Let us denote by
$
\hat{C}_{ij} = J_i T_j +T_i J_j +\frac{1}{2} J^+_iJ^-_j +\frac{1}{2} J^-_iJ^+_j
$,
the Casimir operator acting non-trivially on the $i$ and $j$ factor of
$\otimes_{i=1}^{n} V^{p^+_i,p^-_i}$.
In the coherent state basis this operator can be represented as a differential operator
\be
\hat{C}_{ij} = -p_i^+ p_j^- -p_i^- p_j^+ - (\lambda_j -\lambda_i) (p_i^+ \partial_{\l_j} -p^+_j \partial_{\l_i}), \spa i\neq j~.
\ee
We are going to prove that $G_{\lambda,C}(\vec{z})$ satisfies the KZ equation\footnote{This is true without using any conservation rule!}
\be
\frac{\partial}{\partial z_i}G_{\lambda,C}(\vec{z}) = \sum_{j\neq i}
\frac{\hat{C}_{ij}}{z_i -z_j} G_{\lambda,C}(\vec{z}).
\ee

First it is clear that
\be\label{eqKZ1}
\frac{\partial}{\partial z_i}\prod_{i\neq j} (z_i -z_j)^{-p_i^+p_j^-} =
- \sum_{j\neq i} \frac{p_i^+p_j^- +p_i^-p_j^+}{z_i-z_j} \prod_{i\neq j} (z_i -z_j)^{-p_i^+p_j^-}.
\ee
Then $ \partial_{z_i} \oint_C dw \Delta_{\vec{\l},\vec{p}^+}({w},\vec{z})$ is given by
\be
\oint_C dw \prod_j(w-z_j)^{p_j^+} \frac{p_i^+}{z_i -w}\left(-\sum_j \frac{p_j^+ \l_j}{z_j-w} +
\frac{\l_i}{z_i -w}\right).
\ee
Moreover,
\be
0=\oint_C d\left( \prod_j(w-z_i)^{p_j^+} \frac{p_i^+ \l_i}{z_i -w}\right)
=\oint_C dw \prod_j(w-z_i)^{p_j^+} \frac{p_i^+}{z_i -w}
\left(\sum_j \frac{p_j^+ \l_i}{z_j-w} - \frac{\l_i}{z_i -w}\right).
\ee
Summing the two identities one gets
\bea
\frac{\partial}{\partial z_i}G_{\lambda,C}(\vec{z})
&=&\oint_C dw \prod_j(w-z_i)^{p_j^+} \sum_{j\neq i} \frac{\l_i -\l_j}{z_i-z_j}
\left( \frac{1}{z_j-w} -\frac{1}{z_i -w} \right) \nn
&=& \sum_{j\neq i} \frac{\l_i -\l_j}{z_i-z_j}(p_i^+\partial_{\l_j} -p_j^+ \partial_{\l_i})
\oint_C dw \Delta_{\vec{\l},\vec{p}^+}({w},\vec{z}) ~.
\label{eqKZ2}
\eea
The KZ equation now follows trivially from (\ref{eqKZ1}) and (\ref{eqKZ2}).

\subsection{Flat Space Limit}
\label{sec:flatspace}

There are several unusual features in the amplitude \eqr{corfunc} as we have
written it down which obscure comparison with flat space.
The dependence on H has been made implicit, however
flat space is obtained precisely as H goes to 0.
In particular, what we have implicitly done is to rescale
light-cone momenta according to
\[
p^+ \to {Hp^+}~, \spa p^- \to \frac{p^-}{H}~.
\]
One feature of the amplitude is presence of shifts in the conservation rule of $p^-$,
the shifts are proportional to H and they go away when H is taken to 0.
but the summation over $s$ remains.
The nontrivial issue is due to the remaining integral over
$w$ in \eqr{delta}. When  we reinstate $H$ as above, the $p^+_i$
in the integral get rescaled to zero. Thus the integral reduces simply to
\be
\int d^2w
 \left(  \sum_{i=1}^n \frac{p^+_i\l_i}{w-z_i}    \right)~
\left(  \sum_{j=1}^n \frac{p^+_j\bar{\l}_j}{\bar{w}-\bz_j}    \right)~.
\ee
There are two types of terms appearing when we open the bracket,
depending on whether we combine same or different terms ($i=j$ or $i\neq j$):
\[  \int \frac{d^2 w }{ |w|^2 } ~~~\textrm{and}  ~~~
\int \frac{d^2 w }{ (w-z_i) (\bar{w}- \bz_j) } \]
After properly regularizing, the first one is set to zero,
while the second one is taken to be $-\ln (\mu |z_i - z_j|)$.
Thus the summation over $s$ in \eqr{corfunc} gives
an exponential
\[ \exp ( -\sum_{i \neq j} p^+_ip^+_j\l_i \bar{\l}_j \ln (\mu |z_i - z_j|) ) =
\prod_{i \neq j} |z_i - z_j|^{\frac{-p^i_{\perp}p^j_{\perp}}{4} }~,
\]
if we also set $p_{\perp} = 2p^+ \lambda$.

Then by the usual arguments the final result for the physical amplitude
is shown to be independent of the infrared cutoff $\mu$ as
long as momentum conservation is observed.

\section*{Acknowledgement}
We would like to thank  Curt Callan, Francois Delduc, Jaume Gomis, David Gross,
Juan Maldacena, Chiara Nappi, Rob Myers, Joe Polchinski, Christian
Roemelsberger, Radu Roiban, Yuji Satoh, John Schwarz, Joerg
Teschner and Edward Witten for very useful discussions. We would
like to thank the visitor program of Perimeter Institute for
bringing us together; and KS would like to thank Caltech Theory
Group for extended hospitality. We would also like to thank the
Aspen Center where part of  the work was done. EC was supported in
parts by John McCone Foundation and DOE grant DE-FG03-92-ER40701,
and by the Perimeter Institute's visitor program. LF was supported
in parts by grants UMR 5672 du CNRS and  by the Perimeter
Institute's visitor program. Finally EC and KS would like to thank
Rob Myers and Renee Schingh for support and encouragement.

\appendix
\section{Light Cone Treatment}
\label{app:lc} 

We would now go to the  light-cone gauge so that the model
becomes soluble in this nontrivial background.
The light cone quantization of this model was first studied by
\cite{rt, Forgacs:1995tx}.  The key is that the
cubic interaction term becomes quadratic when one imposes 
$u=X^+ = p^+\tau$. 
In a generic curved background the light cone gauge
cannot be imposed. However it  has been shown by Horowitz and
Steif\cite{HorowitzSteif} that the light-cone choice is consistent 
provided that one has a light-like Killing vector,
as it is the case of the  plane wave background.

we generalize the standard treatment \cite{Callan:2003xr} of light-cone gauge fixing
and the construction of the light-cone Hamiltonian to the case where the NS field
is present.

The action is
\be
        \mathcal{L} = \half \int d^{2} \sigma ~%
        \partial_a X^{m} \, \partial_\beta X^{n}
        \left( \sqrt{h} \, h^{a\beta}  G_{mn} + \epsilon^{a\beta} B_{mn} \right)
\ee
where
\be    \label{app-metricdeterminant}
h= h_{\tau\sigma}\, h_{\tau\sigma} - h_{\tau\tau} \, h_{\sigma\sigma}.
\ee
The generalized momenta are
\be
\Pi_m = \frac{\delta \mathcal{L} }{\delta \dot{X}^{m} }
        = \sqrt{h} \, G_{mn} (h^{\tau\tau} \dot{X}^{n} + h^{\tau\sigma}  \acute{X}^{ n} ) +
                         B_{mn}  \acute{X}^{ n}
\ee
Now solve for $\dot{X}$
\be
 \dot{X}^{m} =\frac{G^{mn}}{\sqrt{h} \, h^{\tau\tau}} \left( \Pi_n - B_{np} \acute{X}^{ p} \right) -\frac{h^{\tau\sigma}}{h^{\tau\tau}} \, \acute{X}^{ m}
\ee
Eliminating $\dot{X}$ from the action, after a tedious computation,
a simple answer emerges:
\be
\mathcal{L} = \half \int d^{2} \sigma  \, \frac{1}{\sqrt{h} h^{\tau\tau} } \, \left( \Pi^2
            - \acute{X}^{ 2} +  \acute{X}^{ m} B_{mn} G^{np} B_{pq} \acute{X}^{ q} \right).
\ee
Note that we have made use of \eqr{app-metricdeterminant} to eliminate $h^{\sigma\sigma}$.
We then Legendre-transform to arrive at the Hamiltonian
\bea 
\label{app-lightconehamiltonian}
\cH &=&\Pi_m \dot{X}^m - \cL  \nn
        &=& \Pi_m \left( \frac{G^{mn}}{\sqrt{h} h^{\tau\tau}} (\Pi_n -B_{np} \acute{X}^{ p})
                -\frac{h^{\tau\sigma}}{h^{\tau\tau}} \acute{X}^{ m} \right) -\cL  \nn
        &=&  \frac{ 1}{2 \, \sqrt{h} \, h^{\tau\tau} }
        \left[   \Pi^2 + \acute{X}^{ 2}
        - 2 \, \Pi_m G^{mn}B_{np} {\acute{X}}^{ p}
         - 2 \acute{X}^{ m} B_{mn} G^{np} B_{pq} \acute{X}^{ q}\right]
          - \frac{ h^{\tau\sigma} }{ h^{\tau\tau} } \, \Pi \cdot \acute{X}^{} \nn
\eea

The independent components of the worldsheet metric above play the role
of Lagrangian multipliers  thus a variation with respect to them
gives the contraint equations.
\bea
\Pi^2 + \acute{X}^{ 2} - 2 \,\Pi^m\, B_{mp} \, \acute{X}^{ p}
-2\acute{X}^{ m} B_{mn} G^{np} B_{pq} \acute{X}^{ q}= 0 \nn
\Pi \cdot \acute{X} = 0
\eea
The first allows us to obtain the light-cone hamiltonian,
while the second expresses the longitudinal coordinate
in terms of transverse physical degrees of freedom.
Light-cone gauge is selected by setting
$X^+ = 2 \pi a^\prime p^+ \tau$ and
$\Pi_- =  2 \pi a^\prime p^+$.

We now specialize to the case of the Nappi-Witten spacetime
with  the metric and NS B-field pertinent to our model,
we get the light-cone constraints
\be \label{app-constraintI} \nonumber
2\,p^+ \Pi_+ + \Pi_1^2 + \Pi_2^2 + 2 \, \mu \, (a_1 \Pi_2 -a_2 \Pi_1)
+ \acute{a}_1^{ 2} +\acute{a}_2^{ 2}
+2 \, \mu \,  (\acute{a}_1  a_2 - \acute{a}_2 a_1) +\mu^2 \, ( a_1^{2} +a_2^{2}) =0
\ee
\be \label{app-constraintII}
\acute{a}_1 \Pi_1 + \acute{a}_2 \Pi_2 + \acute{v} = 0~,
\ee
where $\mu = \frac{p^+H}{2}$. \\

The light-cone Hamiltonian, $\cH_{lc}$, is identified with one of the light-cone momenta:
\bea    \label{app-lightconeH}
\cH_{lc}&\equiv & - p^{+} \Pi_{+}= p^+ \Pi^{-} \nn
2\, \cH_{lc} &=& \Pi_1^2 + \Pi_2^2 + 2 \, \mu \,(a_1 \Pi_2 -a_2 \Pi_1)
   + \acute{a}_1^{ 2} +\acute{a}_2^{ 2} +2 \, \mu \,  (\acute{a}_1 a_2 - \acute{a}_2 a_1)
   +\mu^2 \, ( a_1^{2} +a_2^{2} ). \nn
   &{}&
\eea
And we can also solve for the longitudinal coordinate $\acute{v}$
in terms of the dynamical transverse physical fields:
\bea \label{app-vprime}
\acute{v} &=& -(\acute{a}_1 \Pi_1 + \acute{a}_2 \Pi_2)  \nn
          &=&  - \dot{a_1} a_1^{\prime} - \dot{a_2} a_2^{\prime} -
                                \mu\, ( a_2 a_1^{\prime} - a_1 a_2^{\prime} )~.
\eea
The form of this last expression cannot be naively guessed and differs
from the flat-space result by the H-dependent terms.  This condition,
integrated over $\sigma$, produces the left-right matching
constraint on the physical Hilbert space.

Let us perform a Legendre transformation.  The  light-cone action takes the form:
\be  \label{app-lightconeL}
\cL = \half (\dot{a}_1^{2} +\dot{a}_2^2)  - \half(\acute{a}_1^2 + \acute{a}_2^2)
        - \mu \, (a_1\dot{a}_2- a_2\dot{a}_1  ) + \mu\, (a_1\acute{a}_2-a_2  \acute{a}_1)
\ee
This is naively equivalent to what one gets by setting
$u=X^+ = p^+ \tau$ in the conformal gauge action;
but the subtlety is that in generic curved backgrounds  the conformal gauge is 
incompatible with the light-cone gauge choice. 
For the pp-wave backgrounds this is assured due to existence of null Killing vector
which can be taken as definition of global time\cite{HorowitzSteif}.

\subsection{Solving  the  Equations of Motion}

The equations of motion of the closed string
are, in worldsheet light-cone coordinates $ \sigma^{\pm}=\tau \pm \sigma$,
\be
\partial_+ \partial_{-} a^i + \mu \, \epsilon_{ij} \, \partial_{-} a^j = 0.
\ee
The solutions are
\bea    \label{app-modeexpansion}
a &= (a^1 + i \, a^2)/2
  &= i\, e^{i\mu\sigma^+} \left( \sum_{n\in\mathbb{Z}}
         \frac{\tilde{a}_{n} }{n+\mu} \, e^{-i(n+\mu)\sigma^+}
                     +  \frac{a_{n} }{n-\mu} \, e^{-i(n-\mu)\sigma^-}  \right)
\eea
\bea
\bar{a} &= (a^1 - i \, a^2)/2
  &= i\, e^{-i\mu\sigma^+} \left( \sum_{n\in\mathbb{Z}}
         \frac{\bar{\tilde{a}}_{n}}{n-\mu} \, e^{-i(n-\mu)\sigma^+}
                     +  \frac{\bar{a}_{n}}{n+\mu} \, e^{-i(n+\mu)\sigma^-}  \right)
\eea
With the introduction of a time-dependent twisting,
these can in turn be written in terms of  free fields, $X$, satisfying
$\partial_+\partial_-X=0$, $X(\sigma +2\pi,\tau)=e^{-i2\pi \mu}X(\sigma,\tau)$.:
\bea \label{app-freefieldX}
a &= e^{i\mu\sigma^+} X;\,
\bar{a} &= e^{-i\mu\sigma^+} \bar{X}.
\eea
The free fields $X$ are orbifold fields, but the physical fields must remain periodic.
The corresponding  Lagrangian  in $X$ is the standard free string action.
This is not contradictory to the original interacting theory because the
 field redefinitions are time-dependent.
The choice of the normalization factors in the mode expansion
will be apparent in the next section.

The $\tila_0$ ($\btila_0$) is the center-of-mass coordinates
and should be identified with $\rho$ ($\bar\rho$).
Similarly we should  identify $a_0$ ($\bar{a}_0$) with the radius, 
$\lambda$ ($\bar\lambda$), of the classical trajectory.
For non-zero H, particles on Nappi-Witten space move in circles,
however position and radius of those are arbitrary. One can also see  that
the terms linear in $\tau$ are not allowed and thus there is no zero-mode momentum
operators in the mode expansion above. If one takes the limit
$H \rightarrow 0$, the frequency of the $a_0$ mode goes to zero and
becomes the momentum operator of the limiting flat space.

\subsection{Quantization}
\label{app:quantization}

After solving the classical equations of motion we now proceed to
quantize the system.  We will find that the zero mode operators
become  noncommutative in the background of constant Neveu-Schwarz  flux.

We first compute the (complex) momenta conjugate to $a$ and $\bar{a}$
\[
\Pi=\dot{a} - i \, \mu \, a~,\spa \bar{\Pi}= \dot{\bar{a}} + i \, \mu \, \bar{a}~.
\]
From the oscillator expansion of $a$ \eqr{app-modeexpansion}
we have
\be
\Pi   =   e^{i\mu\sigma^+}
\left[ \sum_{n\in\mathbb{Z}}  \tilde{a}_{n} \,    e^{-i(n+\mu)\sigma^+} + \,
a_n \, e^{-i(n-\mu) \sigma^-}
\right].
\ee
The Poisson bracket
\be
\{ \Pi(\sigma), \ba (\sigma^\prime) \} = \frac{1}{2\pi} \delta(\sigma - \sigma^\prime )
\ee
implies
\bea
i  \{ a_n, \ba_m \} &=&  \half \delta_{n,-m} (m+\mu)         \nn
i   \{ \tila_n, \btila_m \} &=& \half \delta_{n,-m} (m-\mu).
\eea
Replacing  the Poisson brackets with commutators, we finally obtain:
\bea 
	[ a_n , \ba_m ]  &=&   \half \delta_{n,-m}(n-\m),       \nn
 	{[} \tila_{n} , \btila_m {]}  &=& \half \delta_{n,-m}(n+\m).  
\eea
The reality conditions are given by
\be
	a_n^\dagger = \ba_{-n};   \,\,\,   \tila_n^\dagger =\btila_{-n}.
\ee
Recall that we have been working with complexified coordinate
$a$ and $\bar{a}$. We now would like to undo this and write $a_{n} = a^1_n + i \,a^2_n$.
The commutation relations can be safely reproduced by setting
\be
\begin{array}{rclrcl} 
 \displaystyle \left[ a^{1,2}_n, a^{1,2}_m \right] &=&  n \, \delta_{n,-m} \spa & 
\left[ \tila^{1,2}_{n}, \tila^{1,2}_{m}\right] &=&  n \, \delta_{n,-m}\\
 \displaystyle \left[ a^1_n, a^2_m \right]  &=& -i \, \mu \, \delta_{n,-m} \spa & 
\left[ a^1_n, a^2_m \right]  &=& i \, \mu \, \delta_{n,-m} \\
\end{array}
\ee
Unlike in flat space, the same formula holds for the zero mode.
We are forced to make the highly surprising conclusion that the two directions are
noncommutative already at the closed string level:
\[
	[a^1_0 , a^2_0 ] = i \, \mu.
\]
In terms of the complexified coordinates this means that
$ a_0$ and $\bar{\tilde{a}}_0$ are creation operators.

When  the value of $\mu$ is restricted to be $0\leq \mu<1$
creation operators are those with negative indices, $m<0$.
Unlike the case in flat space case, the right
and left moving zero-modes here are not degenerate, 
{\it i.e.} they have frequencies of $+\mu$ and $-\mu$ respectively.
This is reflected in the mode expansion already -- the right and 
left moving zero-modes are independent of each other.

When $\mu=N+\epsilon$ where $N$ is a positive integer the vacuum  is annihilated by
$ a_{N+m}, \bar{\tilde{a}}_{N+m}, \, m>0$ and
$ \tilde{a}_{-N+m}, \bar{{a}}_{-N+m},\, m\geq 0$.
The ``zero modes" are given by $ a_{N}, \tilde{a}_{-N}$ and the corresponding classical
 classical solution is
\be
\frac{ie^{i N\sigma^+}}{\epsilon}
    \left(\tilde{a}_{-N} -a_{N}e^{2i\epsilon\tau} \right).
\ee
When $N=0$ this is the geodesic motion,
 centering at $\tila_0/\mu$ and with a radius $ a_0/\mu$,
oscillating in time at frequency $\mu$.
 But for $N\neq 0$ this describes the motion
of a ``long string,"~%
\footnote{See \cite{Seiberg:1999xz, maloog1} for a general definition and
\cite{kirpio, D'Appollonio:2003dr} for an application of this notion to the Nappi-Witten model.}
\\
\be
 e^{iN\sigma} \left( \frac{\tila_{-N}}{\epsilon} e^{iN\tau} 
 - \frac{a_N}{\epsilon}  e^{iN\tau} e^{2i\epsilon\tau} \right)
\ee \\
{\it i.e.} 
a $2$-dimensional surface winding $N$ times around the origin.
It envelopes the geodesic -- centered at $\tila_{-N}/\epsilon$ and with 
a radius of $a_N/\epsilon$ -- 
and oscillates with a slow frequency $\epsilon$.
What we are witnessing here is a dynamical dielectric effect\cite{Myers:1999ps}
 such that the light-cone momentum is transmuted into winding number under
  the influence of the $H$ field:
For every increase of the light-cone momentum by unit value of 
$\frac{1}{2\pi \alpha\prime^2}H^{-1}$ 
the winding number of the ground state also increases by one.

The quantum Hamiltonian can be obtained by substituting the string mode expansions
into the classical expression and normal-ordering:
\be                    \label{app-ham}
\half \,p^+ \mathcal{H}_{lc} = \sum_{n \in \mathbb{Z}} \left(
:a_{n} \bar{a}_{-n}: \, \frac{n}{n-\mu}+
:\tilde{a}_{n}  \btila_{-n}: \, \frac{n+2\mu}{n+\mu}\right)
+\mu(1-\mu) -\frac{1}{12}.
\ee
The normal ordering constant is given by the identity
$\sum_{n>0}(n+\mu) = -1/12 + \mu(1-\mu)/2$.
One could write the Hamiltonian in a more compact way
 by introducing the properly normalized number operators,
$ N_n = \frac{ a_{n} \bar{a}_{-n } } {n - \mu}$,
$ \tilde{N}_n = \frac{ \tilde{a}_{n} \bar{a}_{-n } } {n + \mu}$.
The total left and right occupation numbers are given by $N=\sum_n nN_n$,  
$\tilde{N}=\sum n\tilde{N}_n$.
The transverse angular momenta are defined to be $J=\sum_n N_n$, $\tilde{J}=\sum_n \tilde{N}_n$.
The light cone  Hamiltonian is then given, up to the normal ordering constant, by
\be
\half p^+ \mathcal{H}_{lc} = N +\tilde{N} + 2\mu J,
\ee
in agreement with Russo-Tseytlin \cite{Russo:2002rq}.
The left-right matching constraint (\ref{app-vprime}),
 becomes formally equivalent to the flat space condition:
\[  N = \tilde{N}~. \]

\section{Nappi-Witten  Algebra, Representations and Wave Functions}
\label{representation}

The Nappi-Witten Lie algebra is a central extension of the two-dimensional Poincare
algebra. The anti-hermitian generators of this algebra are denoted
$J,T,J^1,J^2$.
\be
[J^{+}, J^{-} ] = 2\, i\,  T  \spa  [J, J^{+}] = i\, J^{+} \spa [J,J^{-}]=-i\, J^{-} \spa   [T, \cdot] = 0,
\ee
where $J^{\pm}= J^1 \mp i \, J^2$.
The invariant metric is taken to be
\be
\mathrm{tr}(J \, T)=1, \mathrm{tr}(J^{+}J^{-}) =2,
\ee
which is of signature $-+++$.
This algebra possesses two Casimirs  given by $T$ and $\cC=J^+J^-/2 +J^-J^+/2  + 2\, J\, T$.
It will be important to note that this algebra possesses a linear automorphism $\cP$, first discussed
in \cite{Figueroa-O'Farrill:1999ie}, which acts as a charge conjugation:
\be
\cP({J}^+) =J^-~,~~ \cP({J^-})=J^+~,~~ \cP({T}) =-T~, ~~\cP({J})=-J~.
\ee
A general group element in the Nappi-Witten group can be parametrized by the coordinates
$a_1,a_2,u,v$
\be\label{app:group}
g(a,u,v)= e^{\h aJ_+ +\h \ba J_-} e^{\h uJ +\h vT},
\ee
where $a=(a_1+i \, a_2)/2$.
Note that in terms of the corotating frame coordinates $ x = e^{-i\h u/2}a$ the group element can be written
\be\label{groupdef}
g(a,u,v)=  e^\frac{\h uJ +\h vT}{2} e^{\h xJ_+ +\h \bar{x} J_-} e^\frac{\h uJ +\h vT}{2}.
\ee
The metric on the group is given by
\be\label{ppmet}
 ds^2= \frac{1}{\h^2}\mathrm{tr}(g^{-1}dg )^2 =
 2 \, du \, dv + da_{1}^{2} + da_{2}^{2} -\h ( a_{1} \, da_{2} - a_{2} \,  da_{1} ) \,  du
\ee
From the definition  of $g(a,u,v)$ \eqr{groupdef}  we can compute the product
of two group elements
\be
g_L \,  g_R
= g(a_L+e^{i\h u_L}a_R, u_L+u_R, v_L+v_R+ i\h \, (a_L\bar{a}_R 
e^{-i\h u_L} -a_R\bar{a}_Le^{i\h u_L})),
\ee
where $g_L =g(a_L,u_L,v_L)$ and $g_R=g(a_R,u_R,v_R)$.
We can also compute the inverse
\be
g^{-1}(a,u,v) = g(-ae^{-i\h u}, -u,-v).
\ee
Overall this gives
\be
g_L^{-1}g_R
= g(e^{-i\h u_L}(a_R -a_L), u_R-u_L, v_R-v_L - i\h (a_L\bar{a}_R -a_R\bar{a}_L)).
\ee
The charge conjugation is acting as a parity transformation on the space-time
\be
\cP(g(a,u,v))=g(\bar{a},-u,-v).
\ee
The metric (\ref{ppmet}) is invariant under the isometry group $G_L \times G_R$,
$g\rightarrow g_L^{-1} g g_R$.
Since the generator $T$ is commuting this isometry group is seven dimensional.
When we write the infinitesimal action of this group in terms of the coordinates one gets
\be\label{conscharge2}
\begin{array}{rclrcl}
T_L & = & -\dr_v, &                                  J_R^v & = & \dr_v,\\
J_L &=& -(\dr_u + i(a\dr_a -\bar{a}\dr_{\bar{a}})), & J_R^u &=& \dr_u, \\
J_L^+ & = & -(\dr_{a}  + i \h \ba \dr_v), &          J_R^+ &= & e^{i\h u} (\dr_{a} - i \h\ba  \dr_v ),\\
J_L^- &=& -(\dr_{\ba}  -i \h a \dr_v),    &          J_R^-&=& e^{-i\h u} (\dr_{\ba} +i \h a  \dr_v).
\end{array}
\ee
The generators $T$ generates translation in $v$,
 the generator $J_R -J_L$ and $J_L+J_R$ generate translation in the $u$ direction
and rotation in the transverse plane. The other
generators generate some twisted translations in the transverse
plane. Overall, the symmetry group is 7-dimensional and consists of
two commuting copies  of the Nappi-Witten algebra.

\subsection{Representation Theory} 

The representation theory of this algebra was first discussed in \cite{Kiritsis:jk,kkl}, 
the wave functions were partially discusssed in \cite{kirpio, D'Appollonio:2003dr}.

To construct the unitary irreducible representation of the Nappi-witten  algebra 
we first  identify the operators that commute with all the generators of the algebra.  
There are two such operators, the central generator, $T$, and the quadratic 
Casimir, $\cC$.  The former acts like a scalar in an irreducible representation:
\[     T= ip^+~,      \]
where  $p^+$ being  the light-cone momentum.
The Nappi-Witten algebra admits three types of unitary representations.
The unitary condition  means that
\be
(J^{+})^\dagger= - J^{-};\spa  T^\dagger= - T;\, J^\dagger =-J.
\ee

We now suppose that $p^+ > 0$.
In this case we can define
creation and annihilation operators
\bea
a = \frac{ J^+}{i \sqrt{2p^+}};&\spa&  a^\dagger = \frac{ J^-}{i\sqrt{2p^+}};\\
\left[a,a^\dagger\right] &=&1.
\eea
If we denote by $N=a^\dagger a$ the level operator, we see that
$J+iN$ commutes with everybody.  It is a scalar, which we denote by $-ip^-$.
Once $p^+,p^-$ are fixed the representation is uniquely determined,
we denote this representation by $V^{p^+,p^-}$,
it admits a vaccua $|0,p^+,p^-\ket$ annihilated by
$a \sim J^+$.  We will denote it $|0\ket$ for short, if there is no ambiguity.
The full representation $(|n\ket,\,n\geq 0)$ is obtained by the action 
of $a^\dagger \sim J^-$ on the vaccua, 
\be
|n\ket= {(-iJ^-)^n} |0\ket~.
\ee
The symmetry generators  acting on $V^{p^+,p^-},\,p^+>0$ are given by
\bea
T|n\ket&=& ip^+|n\ket, \\
J^+|n\ket &=& i {2p^+ n } |n-1\ket, \\
J^-|n\ket &=&  i |n+1\ket, \\
J|n\ket &=& i(p^- - n)|n\ket~.
\eea
The scalar product is given by
\be
\bra n|m\ket = (2p^+)^n n! \delta_{n,m}
\ee
and the quadratic Casimir is
\be
\cC=  -2p^+(p^- + 1/2).
\ee
In this representation $-iJ$ admits a highest weight $p^-$.


We can construct in a similar way the conjugate representation
denoted $\widetilde{V}^{p^+,p^-}$.
In this representation 
generators  act as $\cP(J), \cP(T), \cP(J^{\pm})$ with
$J^+$ being a creation operator and $J^-$ an annihilation operator.
$-iJ$ admits a lowest weight given by $-p^-$.  
The symmetry generators act as 
\bea
T|n\ket   &=& -ip^+|n\ket, \\
J^+|n\ket &=& i  |n+1\ket, \\
J^-|n\ket &=& {2i p^+ n } |n-1\ket, \\
J|n\ket   &=& i(-p^- + n)|n\ket~.
\eea
Finally the scalar product between the states is 
\be
\bra n|m\ket = (2p^+)^n n! \delta_{n,m}.
\ee

$V^{p^+,p^-}$ and $\widetilde{V}^{p^+,p^-}$  
share  the same value of the quadratic Casimir.
They are not equivalent but they are related by charge conjugation.
The conjugate representation is dual to the original representation 
in the sense that there exists a nondegenerate invariant pairing
$Q: V^{p^+,p^-}\otimes \widetilde{V}^{p^+,p^-}\rightarrow {\mathbb{C}}$
\be
Q(|n\ket,|m\ket) = (-1)^n (2p^+)^n n! \delta_{n-m}.
\ee

When $p^+=0$ and  $p^-\neq 0$, we can construct \cite{Kiritsis:jk} a series of
representations $V^{0,p^-}_\alpha$   labelled by $p^-$ and a positive number 
$\alpha$ and such that a basis  is given by $|n\ket, n\in Z$.
There are no highest or lowest state
and the action of the symmetry generators is given by
\bea
T|n\ket &=& 0, \\
J^+|n\ket &=& i \alpha|n-1\ket, \\
J^-|n\ket &=& i\alpha |n+1\ket, \\
J |n\ket &=& i(p^- - n)|n\ket,.
\eea
The scalar product is given by
\be
\bra n|m\ket = \delta_{n,m},
\ee
and the Casimir by
\be
\cC=-\alpha^2 .
\ee
These representations do not admit highest or lowest weight states.
Also, are not all inequivalent, if we shift the labels $n$ by one
we get a representation where $p^-$ is also shifted
\be
V^{0,p^-}_\alpha \sim V^{0,p^-+1}_\alpha.
\ee
The last unitary representation is the trivial representation $V^{0,0}$.

\subsection{Coherent States}

For our purpose it is convenient to use the coherent basis representation.
From now on, we consider $p^+>0$, the Hilbert space $V^{p^+,p^-}$.
The coherent state for  $V^{p^+,p^-}$  is constructed as 
\be\label{coherent}
|\lambda\ket = \exp(-\lambda J^-) \, |0\ket,
\ee
in which the symmetry generators are represented by differential operators
\bea\label{cogene}
T|\lambda\ket&=& ip^+~|\lambda\ket, \\
J^+ |\lambda\ket&=& 2p^+ {\lambda}~|\lambda\ket, \\
J^- |\lambda\ket&=& -\partial_{{\lambda}}~|\lambda\ket, \\
J |\lambda\ket&=& i(p^- - {\lambda} \partial_{{\lambda}})~|\lambda\ket,
\eea
The conjugate state is given by
\be\label{conjugate}
\bra\lambda| = \bra 0|\exp(\bar{\lambda} J^+).
\ee
Note that here we are abusing notations, using $|\lambda\ket$ instead of $|\lambda,p^+,p^-\ket$.
The scalar product of such states is given by
\be
\bra\rho|\lambda\ket = e^{2p^+ \bar{\rho} \lambda},
\ee
and the decomposition of unity by
\be
1= \frac{2p^+}{\pi}\int {d^2\lambda} ~e^{-2p^+ \lambda \bar{\lambda}}~
|\lambda\ket \bra\lambda|.
\ee

In the coherent state basis of $\widetilde{V}^{p^+,p^-}$
\be\label{coherentdual}
|\lambda\ket = \exp(-\lambda J^+)|0\ket,
\ee
The symmetry generators are given by
\bea\label{cogenedual}
T|\lambda\ket&=& -ip^+~|\lambda\ket, \\
J^+ |\lambda\ket&=& -\partial_{\lambda}|\lambda\ket, \\
J^- |\lambda\ket&=& 2p^+ {\lambda}~|\lambda\ket, \\
J |\lambda\ket&=& i(-p^- + {\lambda} \partial_{{\lambda}})~|\lambda\ket,
\eea
The scalar product of such states is given by
\be
\bra\rho|\lambda\ket = e^{-2p^+ \bar{\rho} \lambda},
\ee

When $p^+=0$ we define the `coherent state' of the representation
$V^{0,p^-}_\alpha$ to be
\be
|\theta\ket= \sum_{n\in \mathbb{Z}} e^{in\theta} |n\ket
\ee
The symmetry generators are given by
\bea\label{cogene0}
T|\theta\ket&=& -ip^+~|\theta\ket, \\
J^+ |\theta\ket&=& i \alpha e^{i\theta}|\theta\ket, \\
J^- |\theta\ket&=& i \alpha e^{-i\theta}|\theta\ket, \\
J |\theta\ket&=& i(p^- - \partial_{{\theta}})~|\lambda\ket,
\eea
and the scalar product by
\be
\bra\phi|\theta\ket = \delta(\phi -\theta).
\ee

\subsection{Wave Functional}

As we have already seen a Nappi-Witten group element can be parametrized by the coordinates
$a,\bar{a},u,v$
\be
g(a,u,v)= e^{\h aJ^+ +\h \ba J^-} e^{\h uJ +\h vT}
\ee
where $a=a_1+i \, a_2$.
We want to compute the matrix element of $g$ in the coherent state basis of
$V^{p^+,p^-}$, let us denote
\be\label{app-defwave}
\phi^{p^+,p^-}_{\bar{\rho},\lambda}(g)= {\bra\rho|g|\lambda\ket}~.
\ee
With the help of the formulae (\ref{coherent}, \ref{app:group}) and after some algebra
where one pushes the $J^+$ to the right and the $J^-$ to the left, one gets
\be\label{wavefuncapp}
\phi^{p^+,p^-}_{\bar{\rho},\lambda}(g)
= e^{ip^+v +ip^-u}
e^{-p^+ a\bar{a}}
\exp\left[2 p^+(a\lambda e^{-iu} -\bar{\rho} \bar{a}) + 2 p^+ \bar{\rho} \lambda e^{-iu}\right],
\ee
The ground state wave functional is
\be
\phi^{p^+,p^-}_{0,0}(g)
= e^{ip^+v +ip^-u} e^{-p^+ a\bar{a}},
\ee
This corresponds to a plane wave centered in the transverse plane around $a = 0$.
In general this wave functional can be written in a more suggestive form as
\bea
\phi^{p^+,p^-}_{\bar{\rho},\lambda}(a,u,v)
&=& e^{ip^+v +ip^-u} \\
&\times& e^{-p^+|\bar{a}-\lambda e^{-iu} +\rho|^2}
e^{p^+a (\rho + \lambda e^{-iu}) - c.c} \\
&\times & e^{p^+|\lambda|^2 + p^+ |\rho|^2 - p^+( \bar{\lambda}\rho e^{iu} - \lambda \bar{\rho} e^{-iu})}.
\eea
This expression clearly shows that the wave functional is centered around
$\bar{a} = -\rho+\lambda e^{-iu}$. Moreover, it is a plane wave in $v$ of momentum $p^+$ and the semi-classical
momentum $p^a$ in the transverse plane is given by $ \rho +\lambda e^{iu}$.
The direction $u$ has a more complicated semi-classical momentum given by
$p^u = p^- - \lambda(a+\bar{\rho})e^{-iu} + c.c$.
All these results agree  with the analysis of the geodesics in Nappi-Witten space.

Using the same techniques we can construct the matrix elements of $g$ in the conjugate  representation
\be
\widetilde{\phi}^{p^+,p^-}_{\bar{\rho},\lambda}(g)\equiv
{\phi}^{p^+,p^-}_{\bar{\rho},\lambda}(\cP(g))
= e^{-ip^+v - ip^-u} e^{-p^+ a\bar{a}}
\exp\left[2 p^+(\bar{a}\lambda e^{+iu} -\bar{\rho}{a}) + 2 p^+ \bar{\rho} \lambda e^{+iu}\right]~.
\ee
This corresponds to a wave moving in backwards in light-cone time and centered around  $a= \lambda e^{iu} -{\rho}$.
The conjugate wave functional is related to the original one by complex conjugation
\be
\widetilde{\phi}^{p^+,p^-}_{\bar{\rho},\lambda}(g)=
\overline{{\phi}^{p^+,p^-}_{{\rho},\bar{\lambda}}(g)}.
\ee

One can finally construct the wave functional associated with the $p^+=0$ representations.
The matrix elements of $g$ in the representation $V^{0,p^-}_\alpha$ are given by
\be
\bra\phi|g|\theta\ket= \delta(\phi-\theta +u)
e^{ip^-u} e^{ i\alpha e^{i\theta} e^{-iu} a  +i \alpha e^{-i\theta} e^{iu}  \bar{a}}.
\ee

\subsection{Wave Equation}

The wave functions $\phi^{p^+,p^-}_{\bar{\rho},\lambda}$ are a complete basis of
solutions of the wave equations
\be \label{app:waveeq}
\partial_v \phi = ip^+ \phi, \spa
\Delta \phi = -2p^+(p^- -1/2)\phi,
\ee
where $p^+>0$ and
\be
 \Delta = 2 \partial_u \partial_v +\partial_a \partial_{\bar{a}} + a \bar{a} \partial_v \partial_v
+ i(a\partial_{a} -\ba \partial_{\ba}) \partial_v,
\ee
is the Laplacian of the pp-wave metric.
This Laplacian is also equal to the Casimir for the left or right symmetry generators (\ref{conscharge2})
\bea
\Delta &=& 2J_L T_L + (J^+_LJ^-_L +J^-_LJ^+_L)/2,\\
&=&  2J_R T_R + (J^+_RJ^-_R +J^-_RJ^+_R)/2.
\eea
This can be checked directly but can also be understood by the fact that the wave
function $\phi^{p^+,p^-}_{\bar{\rho},\lambda}$ intertwines the action
of the left symmetry generators with the coherent state generators (\ref{cogene})
acting on $\bar{\rho}$.
One can check that $\phi^{p^+,p^-}_{\bar{\rho},\lambda}(g)$ is a solution of
\be\label{symaction}
\begin{array}{rclrcl}
T_L \phi(g)    &=&-ip^+ \phi(g),&\spa T_R \phi(g)    &=& ip^+ \phi(g),\\
J_L^+ \phi(g) &=& -\partial_{\bar{\rho}}\phi(g),& \spa  J_R^+ \phi(g) &=& 2p^+\lambda \,\phi(g)\\
J_L^- \phi(g) &=&  2 p^+\bar{\rho} \,\phi(g),& \spa J_R^- \phi(g) &=& -\partial_{\lambda}\phi(g) \\
J_L \phi(g) &=& -i(p^- - \bar{\rho}\partial_{\bar{\rho}})\phi(g),&\spa J_R \phi(g) &=&
i(p^- - \lambda\partial_\lambda)\phi(g)~.
\end{array}
\ee
This is clear by the construction of the wave function since
\be
J_R^+ \bra\rho|g|\lambda\ket =\bra\rho| g J^+|\lambda\ket= 2p^+\lambda \bra\rho| g|\lambda\ket,
\ee
and similarly for the other generators.

The wave functions $\widetilde{\phi}^{p^+,p^-}_{\bar{\rho},\lambda}$ are
solutions of the wave equations
\be
\partial_v \phi = -ip^+ \phi; \spa
\Delta \phi = -2p^+(p^- +1/2)\phi~.
\ee
The action of the symmetry generators on $\widetilde{\phi}^{p^+,p^-}_{\bar{\rho},\lambda}$
is obtained from (\ref{symaction}) by exchanging the left symmetry generators with the right generators.

It is important for us to note that $\phi^{-p^+,-(p^-+1)}_{\bar{\rho},\lambda}(g)$
is also solution of \eqr{app:waveeq}.
This solution is well defined as a function on the group, however
it contains a factor $ \exp(2p^+ a\bar{a})$ which is unbounded.
It is then not normalisable and can not be taken as a wave functional corresponding
to the representation  $\widetilde{V}^{p^+,p^-}$.
However one can check  that the integral transform of this solution
\be\label{duality2}
\widehat{\phi}^{p^+,p^-}_{\bar{\rho},\rho}(g) \equiv
\int  e^{2p^+\l \rho} e^{2p^+\bar{\l} \bar{\rho}}\,{\phi}^{-p^+,-(p^-+1)}_{\bar{\l} \l}(g)\,d\l d\bar{\l},
\ee
satisfies the same transformation rules as $\widetilde{\phi}^{p^+,p^-}_{\bar{\rho},\rho}$
 under the symmetry transformation.
 In order to show this we only need  to use   rule of integration by part
 \be
 \widehat{\phi}^{p^+,p^-}_{\rho,\lambda}(g) =
 \widehat{\phi}^{p^+,p^-}_{\rho,\lambda}(1) \widetilde{\phi}^{p^+,p^-}_{\rho,\lambda}(g).
\ee
This relation is formal\footnote{The proportionality coefficient between $\widehat{\phi}$ and
$\widetilde{\phi}$ could be infinite.} if we do not check the convergence properties of the integral transform.
Ideally we would have to specify a proper choice of contour of integration in order for this relation
to be rigourously valid. In Appendix~\ref{Itransf} we give the rules for computing such integral transforms.

\subsection{Plancherel formula}
We denote by $\chi^{p^+,p^-}(g) ={\rm{tr}}_{V^{p^+,p^-}}(g)$ the character of the representation
$V^{p^+,p^-}$, $\widetilde{\chi}^{p^+,p^-}(g)$ and $\chi^{0,p^+}_\alpha(g)$
the one associated with $\widetilde{V}^{p^+,p^-}$ and $V^{0,p^-}_\alpha$.
A direct computation using $\chi^{p^+,p^-}(g) =\frac{2p^+}{\pi} \int e^{-2p^+ \l \bar{\l}} \,
\phi_{\bar{\l} \l}^{p^+,p^-}(g) d^2\l $
 shows that
\be
\chi^{p^+,p^-}(g) =  \frac{e^{ip^+ v} e^{ip^- u}}{1-e^{-iu}}\exp\left( {i}{p^+} \frac{\cos u/2}{\sin{u/2}}  a \bar{a}\right)
\ee
when $ u \neq 0 (2\pi)$.
If $u= 0(2\pi)$ then $\chi^{p^+,p^-}(g) = \frac{\pi}{2p^+} \delta^2(a)$.
Also,
$\widetilde{\chi}^{p^+,p^-}(g)=  \overline{\chi^{p^+,p^-}(g)}$
and
\be
\chi^{0,p^-}_\alpha(g)= \delta(u) J_0(2\alpha |a|)
\ee
where $J_0$ is the Bessel function.
Now
\be
\int_{-\infty}^{+\infty}dp^- \chi^{p^+,p^-}(g)=
\frac{\pi^2}{p^+} e^{ip^+ v} \delta(u) \delta^2(a).
\ee
We therefore have the Plancherel formula
\be
\frac{1}{\pi^2} \int_0^{+\infty} p^+ d{p^+}\int_{-\infty}^{+\infty} dp^-
(\chi^{p^+,p^-}(g)+\widetilde{\chi}^{p^+,p^-}(g)) = \delta(g).
\ee

\subsection{More Representation Theory}

The way the product of two waves functions decomposes as a linear sum of
wave function gives us  a lot of important information from the physical and mathematical side.
From the mathematical side it contains all the information we need on the tensor product of
representation and the recoupling coefficient involved in the tensorisation.
From the physical side we can read out what are the conservation rules and how two wave interact
in our curved background.
We are interested in the multiplicative properties of the wave
functions
\bea\label{waveprod}
\phi^{p^+,p^-}_{\bar{\rho},\lambda}(g)
&=& e^{ip^+(v+ia\bar{a}) +ip^-u} e^{2p^+ a\lambda e^{-iu}} e^{-2p^+ \bar{\rho} \bar{a}}
e^{{2p^+}{\bar{\rho}}\lambda e^{-iu}}\\
\widetilde{\phi}^{p^+,p^-}_{\bar{\rho},\lambda}(g)
&=& e^{-ip^+(v -ia\bar{a}) -ip^-u} e^{2p^+\bar{a}\lambda e^{+iu}} e^{-2p^+\bar{\rho} {a}}
e^{{2p^+}{\bar{\rho}}\lambda e^{+iu}}~.
\eea
It is easy to see that if we define
$p^\pm_3 =p^\pm_1 +p^\pm_2 $, $ p^+_3 \l_3 = p^+_1\l_1 + p^+_2\l_2 $, $p^+_3\rho_3=p^+_1\rho_1+p^+_2\rho_2$
the product is given by
\be
\phi^{p^+_1,p^-_1}_{\bar{\rho}_1,\lambda_1}(g)\phi^{p^+_2,p^-_2}_{\bar{\rho}_2,\lambda_2}(g)=
\phi^{p^+_3,p^-_3}_{\bar{\rho}_3,\lambda_3}(g)
\exp e^{-iu}({2p^+_1}{\bar{\rho_1}\lambda_1} +{2p^+_2}{\bar{\rho_2}\lambda_2}-
{2p^+_3} {\bar{\rho_3}\lambda_3}).
\ee
The term in the exponent can be evaluated to be
\be
\frac{p_1^+p_2^+}{p_3^+}({\l_1} -{\l_2})
({\bar{\rho}_1}-{\bar{\rho}_2})e^{-iu}.
\ee
If we Taylor expand the exponential and
utilize the fact that
$e^{-iu} \phi^{p^+,p^-}_{\bar{\rho},\lambda} =
 \phi^{p^+,p^- - 1}_{\bar{\rho},\lambda}$,
 we get the simple result
 \be
\phi^{p^+_1,p^-_1}_{\bar{\rho}_1,\lambda_1}(g)\phi^{p^+_2,p^-_2}_{\bar{\rho}_2,\lambda_2}(g)
=
\sum_{n=0}^{\infty}
\frac{1}{n!}\left(\frac{p_1^+p_2^+}{p_3^+}\right)^n
({\l_1} -{\l_2})^n
({\bar{\rho}_1}-{\bar{\rho}_2})^n
\phi^{p^+_3,p^-_3-n}_{\bar{\rho}_3,\lambda_3}(g).
\ee
From which we read out the tensorisation rules
\be
V^{p^+_1,p^-_1} \otimes V^{p^+_2,p^-_2} =
\sum_{n=0}^{+\infty} V^{p^+_1 +p^+_2,p^-_1+p^-_2 -n},
\ee
and the Clebsh-Gordan coefficients
\be
C_{\l_1 \l_2 p^\pm_3}^{p^\pm_1 p^\pm_2 \l_3} =
\sum_n \delta( p_1^+ + p^+_2 -p^+_3)\delta(p_1^- +p^-_2-p^-_3 -n)\delta^2(\l_1+\l_2 -\l_3)
\frac{1}{\sqrt{n!}}\left(\frac{p_1^+p_2^+}{p_3^+ }\right)^\frac{n}{2}
({\l_1} -{\l_2})^n~.
\ee
The product of two conjugate wave function is
similar since $\widetilde{\phi}(g) = \phi(\cP(g))$ and so is the tensorisation rule and the
Clebsch-Gordan coefficients  for the conjugate representations $\widetilde{V}$.

\section{Integral Transform}
\label{Itransf}
We define the integral transform to be the following linear transformation
\be
I(\l^s \bar{\l}^{\bar{s}})(\rho,\bar{\rho}) =
\frac{\Gamma(\bar{s}+1)}{\Gamma(-s)}
(-\bar{\rho})^{-\bar{s}-1}({\rho})^{-{s}-1},
\ee
with $s-\bar{s}$ restricted to be an integer.  Under this condition
this expression is symmetric in the exchange of $s$ with $\bar{s}$.
One can check that it satisfies the following properties
\bea\label{intid}
I(e^{ \bar{a} \bar{\l}}(\l \bar{\l})^s)(\rho,\bar{\rho}) &=&
I((\l \bar{\l})^s)(\rho  ,\bar{\rho} +\bar{a})\\
I((\l+a)^s \bar{\l}^{\bar{s}})(\rho) &=& e^{-\rho a}I(\l^s \bar{\l}^{\bar{s}})(\rho)\\
I\circ I((\l^s \bar{\l}^{\bar{s}})(\rho)&=& -(-1)^{s-\bar{s}} \rho^s \bar{\rho}^{\bar{s}}
\eea
the first two identities express the fact that $I$ represent the integral transform
\be
I(f)(\rho,\bar{\rho})= \frac{-1}{\pi} \int d^2\l |e^{\rho \l}|^2 f(\l,\bar{\l})
\ee
where the normalization has been fixed by
computing the integral transform of $e^{\lambda \bar{\lambda}}$.

In order to prove these equalities we need to use the following identities
\bea    \label{int+}
e^{-z} &=&\sum_n \frac{(-1)^n}{\Gamma(n+1)} z^{n}, \\
(z+w)^{s}&=& \sum_n \frac{(-1)^n \Gamma(n-s)}{\Gamma(n+1)\Gamma(-s)} z^n w^{n-x}
\eea
when $z<w$.
Let us prove for instance the first identity \eqr{intid}
\bea 
I(e^{ \bar{a} \bar{\l}}(\l \bar{\l})^s)(\rho,\bar{\rho}) 
&=&\sum_n \displaystyle\frac{\bar{a}^n}{\Gamma(n+1)} I( \l^n \bar{\l}^{n+s}) \nn 
&=&\sum_n \displaystyle\frac{\Gamma(n+s+1)}{\Gamma(n+1) \Gamma(-s)} \bar{a}^n \rho^{-s-1} (-\bar{\rho})^{-n-s-1}  \nn
&=&\displaystyle\frac{\Gamma(s+1)}{\Gamma(-s)} (-\bar{\rho}\rho)^{-s-1}
\sum_n \frac{\Gamma(n+s+1)}{\Gamma(n+1)\Gamma(s+1)}  \left(-\frac{\bar{a}}{\bar{\rho}}\right)^n \nn
&=&\displaystyle\frac{\Gamma(s+1)}{\Gamma(-s)}(-\rho (\bar{\rho}+\bar{a}))^{-s-1}
\eea
which is the RHS of \eqr{intid}.
The main result of this section consists in showing the following identity
\be\label{mainres}
I\left( \frac{(A\bar{\l}\l +\bar{B} \l +B \bar{\l} +C)^s}{\Gamma(s+1)}\right)(\rho,\bar{\rho})
= -\left|e^{-\rho \frac{B}{A}}\right|^2 \frac{ A^s}{(-|\rho|^2)^{s+1}}
\left(\frac{\cal{D}}{2}\right)^{s+1}  I_{-s-1}({\cal{D}})
\ee
where we have introduced
\be
\frac{\cal{D}}{2} = |\rho| \frac{\sqrt{B\bar{B} -AC}}{A},
\ee
and $I_\nu(z)$ is the modified Bessel function of the first kind.

Here is the proof: first we complete the square
\be
(A\bar{\l}\l +\bar{B} \l +B \bar{\l} +C)
= A \left( \left|\l + \frac{B}{A}\right|^2 + \frac{AC-B \bar{B}}{A^2}\right)
\ee
then use the identity \eqr{int+} to express the LHS of (\ref{mainres}) as
\be
\sum_n \frac{(-1)^n \Gamma(n -s)}{\Gamma(n+1) \Gamma(-s)\Gamma(s+1)} A^s
I\left(\left|\l + \frac{B}{A}\right|^{2(s-n)}\right)(\rho) \left(\frac{AC-B \bar{B}}{A^2}\right)^{n}.
\ee
after evaluation of the integral transform one get
\be
\left|e^{-\rho \frac{B}{A}}\right|^2 \frac{A^s}{(-|\rho|^2)^{s+1}}\frac{\sin\pi(s+1)}{\pi}
\sum_n \frac{(-1)^n\Gamma(s+1 -n)}{\Gamma(n+1)}
\left(-|\rho|^2 \frac{AC-B \bar{B}}{A^2}\right)^{n}.
\ee
To finish the proof one need to use the identity
\be
\sum_n \frac{(-1)^n \Gamma(\nu-n)}{\Gamma(n+1) } \left(\frac{z}{2}\right)^{2n}
=-\frac{\pi}{\sin\pi\nu}\left(\frac{z}{2}\right)^{\nu}  I_{-\nu}(z)
\ee

\section{Chiral Splitting of Integrals}
\label{app:split}

In this section we prove  the chiral splitting property  of the integral
\be \label{split}
I(f,\bar{g})= \int d^2w \, f(w) \, \bar{g}(\bar w)
\ee
where
$f,\bar{g}$ are analytic functions having branch points at $z_1,\cdots, z_n$.  The monodromy around the point $z_i$ is given by   $e^{2i\pi p_i}$.
For example  we can have $ f(w)= \prod_i(w-z_i)^{p_i}$.
The main point  we want to stress here is that such surface integrals can be written in terms of  contour integrals.
Suppose for simplicity that all $z_i$ are real and ordered as $ z_1< z_2<...<z_n$.
We then have
\be
\int d^2w ~f(w) \, \bar{g}(w)
= \sum_{i=1}^{n-1} \oint_{a_i} f(w) \, dw \, {\int_{z_i}^{z_{i+1}} \bar{g}(\bar w) d\bar w}~,
\ee
where $a_i$ is the contour of integration starting at $-\infty$ going around~$z_1,\cdots,z_i$
 in a clockwise direction and going back to $-\infty $ below the real axis.
The proof of this statement is obtained by writing the integral over $w=x+iy$
as an integral over $x$ and $y$ and then rotating the contour of integration of $iy$ along the real axis.
A careful analysis similar to the one done in \cite{Kawai:1985xq} leads to the results stated above.

We can now specify this result  when there are three finite branch points
$0,z,1$ with $0<z<1$.
Let us introduce the following notation
$ I_1(f) =\int_{-\infty}^0 |f(w)|dw $,  $ I_2(f) =\int_0^z |f(w)|dw $, $I_3(f) =\int_{z}^1 |f(w)|dw $,
$I_4(f) =\int_{1}^\infty |f(w)|dw $.
These integrals are not all independent.  Consider the contour of integration $C$
starting from $+\infty$ going around $0,z,1$ in a clockwise manner and going back to $\infty$.
The integral of $f$ along this contour is zero; and it can be decomposed as integrals $I_i$.
Namely
\be\label{id1}
0= \oint_C f = \sin{\pi p_1} I_2 + \sin{\pi(p_1 +p_2)} I_3 +\sin\pi(p_1+p_2+p_3)I_4
\ee
If we now take a contour which starts at $-\infty$, goes around $0,z$ in a clockwise direction, comes back at
$-\infty$  and then  another contour which starts off at $+\infty$ and goes around 1 in a counterclockwise direction, we get another identity
\be\label{id2}
0=\sin{\pi (p_1+p_2)} I_1 + \sin{\pi p_2} I_2 - \sin\pi p_3 I_4.
\ee
The integrals along the contours $a_1,a_2$ can also be expressed in terms of $I_i$
\bea
\oint_{a_1} \frac{d^2\lambda}{\pi}  f = - \sin\pi p_1 I_1(f) \\
\oint_{a_2} \frac{d^2\lambda}{\pi}  f = - \sin\pi p_3 I_4(f).
\eea
The measure in the above integrals are fixed by
computing the integral transform of $e^{\lambda \bar{\lambda}}$.
Altogether the total integral can be written as
\be
- \sin\pi p_1 I_1(f)\bar{I}_2(g)- \sin\pi p_3 I_4(f)  \bar{I}_3(g),
\ee
Using \eqr{id1} and \eqr{id2}  we can eliminate $I_1(f)$ and $\bar{I}_3(g)$ respectively.
This finally gives the identity
\be\label{finalchirsplit}
I(f,\bar{g})=\frac{\sin\pi p_1 \sin\pi p_2}{\sin \pi (p_1 +p_2)}I_2(f)\bar{I}_2(g)
+\frac{\sin\pi p_3 \sin\pi (p_1+p_2+p_3)}{\sin \pi (p_1 +p_2)} I_4(f)\bar{I}_4(g)~.
\ee

\addcontentsline{toc}{section}{References}
\bibliographystyle{JHEP}

\end{document}